\documentclass[a4paper,11pt]{article}
\usepackage[a4paper,top=2.5cm,bottom=2.8cm,left=2.6cm,right=2.6cm]{geometry}
\usepackage[T1]{fontenc}
\usepackage{lmodern}
\usepackage{microtype}
\usepackage{amsmath,amssymb}
\usepackage{mathtools}
\usepackage{bm}
\usepackage{graphicx}
\usepackage{booktabs}
\usepackage{array}
\usepackage{enumitem}
\setlist{itemsep=1pt,topsep=3pt,parsep=0pt}
\usepackage[font=small,labelfont=bf]{caption}
\usepackage[siunitx]{circuitikz}
\usepackage[numbers,sort&compress]{natbib}
\usepackage{authblk}
\usepackage[colorlinks=true,linkcolor=blue,citecolor=blue,urlcolor=blue]{hyperref}

\numberwithin{equation}{section}

\newcommand{\PT}{\mathcal{P}\mathcal{T}}
\newcommand{\Pex}{\mathcal{P}}
\newcommand{\Trev}{\mathcal{T}}
\newcommand{\dd}{\mathrm{d}}
\newcommand{\pcm}{p_{\mathrm{cm}}}
\newcommand{\Uop}{\mathcal{U}}
\newcommand{\E}{\mathcal{E}}
\newcommand{\Gen}{\mathcal{G}}
\IfFileExists{dsfont.sty}{\usepackage{dsfont}\newcommand{\one}{\mathds{1}}}{\newcommand{\one}{\mathbb{I}}}

\newcommand{\figfile}[2]{\IfFileExists{#2}{\includegraphics[width=#1]{#2}}{\fbox{\parbox[c][4cm][c]{0.8\linewidth}{\centering Figure file \texttt{\detokenize{#2}} to be added}}}}

\title{Time-reflection nonlocal equations from $\PT$-symmetric oscillator pairs: physical origin and variational formulation}

\author[1,2]{Ilmar Gahramanov\thanks{\href{mailto:ilmar.gahramanov@bogazici.edu.tr}{ilmar.gahramanov@bogazici.edu.tr}}}
\author[1]{Ya\u{g}\i zhan \c{S}ahin\thanks{\href{mailto:yagizhansahin37@gmail.com}{yagizhansahin37@gmail.com}}}
\affil[1]{Department of Physics, Bo\u{g}azi\c{c}i University, 34342 Bebek, Istanbul, T\"urkiye}
\affil[2]{Center for Mathematics and its Applications, Khazar University, Mehseti St.~41, AZ1096, Baku, Azerbaijan}
\date{}

\begin{document}
\maketitle

\begin{abstract}
We show that every second-order differential equation with time reflection describes the $\PT$-symmetric solutions of a local pair of oscillators that is invariant under exchange combined with time reversal. For linear equations the pair is built from springs, balanced gain and loss, and a gyroscopic coupling. We determine when such a pair has a regular first-order Lagrangian, and when the solutions of the equation fix its coefficients. As the main example, we consider a recently studied equation whose solutions do not depend on its velocity coupling. In the corresponding pair this coupling acts only on the conserved total momentum, which vanishes for $\PT$-symmetric solutions, and each of the two conditions that single out the model places the pair at an exceptional point. When the spring and velocity couplings are both nonzero, the pair has no regular first-order Lagrangian, while the equation has one with second derivatives. We also propose an electronic realization and construct a Bateman-type Hamiltonian, from which the velocity coupling is removed by a canonical transformation.
\end{abstract}


%======================================================================
\section{Introduction}\label{sec:intro}

Differential equations in which the unknown function is evaluated at a reflected argument form a classical subclass of functional differential equations~\cite{Wiener1969,PrzeworskaRolewicz1973,WienerAftabizadeh1985,Wiener1993,CabadaTojo2015}. The standard method for them is also classical: the equation is reduced to ordinary differential equations, by differentiation or for the even and odd parts of the solution~\cite{WienerAftabizadeh1985,CabadaTojo2015}. This reduction was recently used to write down the general solutions of first-order and second-order linear equations of this type~\cite{Ma2024,Mandal2025}, and of a class of linear equations of higher order~\cite{Yuan2025}. Related recent work treats reflection equations with piecewise constant arguments~\cite{CabadaCambeses2026} and first-order reflected equations with polynomial forcing~\cite{Albidah2026}. The physical origin of these equations, however, is rarely discussed. $\PT$-symmetric systems are mentioned as an application in~\cite{Mandal2025}. Here we derive these equations from such systems.

In physics, reflected arguments appear in nonlocal integrable equations, starting with the reverse-space nonlinear Schr\"odinger (NLS) equation, which is obtained from a local system by a symmetry reduction~\cite{AblowitzMusslimani2013}; see~\cite{AblowitzMusslimani2017,Lou2018} for reverse-time versions and Alice--Bob systems. A Lagrangian and Hamiltonian formulation of spatially nonlocal equations of this type was recently given in~\cite{Pazarci2026}. In~\cite{Yang2018} nonlocal NLS equations were obtained from the Manakov system, which is invariant under the exchange of its two components combined with a spatial reflection, or with time reversal and complex conjugation, $v(x,t)=u^*(x,-t)$: for symmetric initial data each component obeys a nonlocal equation. Without conjugation, the reductions $r(x,t)=\pm q(x,-t)$ of the AKNS system give the reverse-time NLS equations of~\cite{AblowitzMusslimani2017}, where equations nonlocal in time are understood as solutions of a local evolution system with related initial data.

Systems with balanced gain and loss are studied in the framework of parity--time ($\PT$) symmetry~\cite{BenderBoettcher1998,ElGanainy2018}. A pair of coupled oscillators, one with gain and one with loss of the same rate, has a Hamiltonian and shows $\PT$ phase transitions~\cite{Bender2013PRA}. Such pairs have been realized with active $LRC$ circuits~\cite{Schindler2011,Schindler2012}, coupled pendula~\cite{Bender2013AJP} and whispering-gallery microcavities~\cite{Peng2014}. Their Hamiltonian goes back to Bateman's doubling of the damped oscillator~\cite{Bateman1931,FeshbachTikochinsky1977,Celeghini1992,BlasoneJizba2004}.

A related model of Newtonian type, $\ddot x(t)=\alpha x(t)+\beta x(-t)+f(t)$ in our notation, was solved in~\cite{Ma2024Pramana} by splitting the solution into its even and odd parts. In~\cite{Ma2025} the following equation was considered (see also~\cite{Mandal2025}):
\begin{equation}
  \ddot x(t) = \lambda\bigl[x(t)-x(-t)\bigr] + \mu\bigl[\dot x(t)-\dot x(-t)\bigr],
  \qquad \lambda,\mu\in\mathbb{R}.
  \label{eq:model}
\end{equation}
Here $\dot x(-t)$ denotes the derivative of $x$ evaluated at $-t$, so that $\dd[x(-t)]/\dd t=-\dot x(-t)$, and $x$ is twice differentiable on $\mathbb{R}$. A striking feature of this equation, observed in~\cite{Ma2025}, is that its general solution does not depend on $\mu$.

Equation~\eqref{eq:model} is nonlocal in time: the acceleration at time $t$ depends on the state at time $-t$. So it does not look like the equation of motion of a physical system. In fact every second-order equation of the form $\ddot x(t)=h\bigl(x(t),x(-t),\dot x(t),\dot x(-t)\bigr)$ describes the $\PT$-symmetric solutions of a local pair of oscillators that is invariant under exchange combined with time reversal (Section~\ref{sec:reduction}). For~\eqref{eq:model} this pair is particularly simple: a single spring, balanced gain and loss, and a gyroscopic coupling of the same strength, which can be realized as an electronic circuit. The nonlocality only reflects a symmetry of this pair together with symmetric initial data.

Equation~\eqref{eq:model} is also not variational in an obvious way. Its operator $\E$ is not self-adjoint, $\E-\E^\dagger=-2\mu D$ (Section~\ref{sec:notation}): the reflected term $\mu\dot x(-t)$ is self-adjoint, while the local term $\mu\dot x(t)$ is not. Since the operator of a linear Euler--Lagrange equation is self-adjoint~\cite{Douglas1941,Santilli1978}, for $\mu\neq0$ equation~\eqref{eq:model} is not itself the Euler--Lagrange equation of a Lagrangian depending on $x$ alone. The damped harmonic oscillator has the same problem, and a Lagrangian is obtained either with the multiplier $e^{-\mu t}$~\cite{Bateman1931,Kanai1948}\footnote{For the damped oscillator $\ddot x=\mu\dot x-\omega^2x$, with $\mu<0$, the Lagrangian is $e^{-\mu t}\bigl(\frac12\dot x^2-\frac12\omega^2x^2\bigr)$.} or with an auxiliary degree of freedom~\cite{Bateman1931}. Since the pair is the physical system, we ask for a Lagrangian that describes its full dynamics, not only its $\PT$-symmetric solutions.

Beyond the classical even/odd reduction and the independence of $\mu$ observed in~\cite{Ma2025}, our main results are the following. The first three hold for the general equation with reflection. The others concern the model~\eqref{eq:model}, which we use as the main example.
\begin{itemize}
\item \emph{Reduction to a local pair (Section~\ref{sec:origin}).} Every equation $\ddot x(t)=h\bigl(x(t),x(-t),\dot x(t),\dot x(-t)\bigr)$ describes the $\PT$-symmetric solutions of a local pair of oscillators. For the linear equation the pair contains only springs, balanced gain and loss, and a gyroscopic coupling.
\item \emph{Lagrangians of the pair (Section~\ref{sec:variational}).} The linear pair has a regular first-order Lagrangian unless the two oscillators are coupled by a spring and gain and loss and the gyroscopic coupling have the same strength. In this case it has none, with any multiplier. In Douglas's classification~\cite{Douglas1941} this is his Case~IIIb, and we give a direct derivation in terms of the pair.
\item \emph{Solutions and coefficients (Section~\ref{sec:family}).} The solutions of the general linear equation with reflection determine its coefficients unless $a+b=0$. On this surface there is a one-parameter redundancy.
\item \emph{The model: similarity and physical origin (Sections~\ref{sec:model} and~\ref{sec:origin}).} Equation~\eqref{eq:model} is obtained from its $\mu=0$ version by a similarity transformation $\Uop$ that differs from the identity by a nilpotent operator. In its pair $\mu$ multiplies the conserved total momentum. We give an electronic realization.
\item \emph{The model: Jordan blocks (Section~\ref{sec:spectral}).} Among the linear pairs, the model is singled out by two conditions: gain and loss and the gyroscopic coupling have the same strength, and there is no on-site spring. Each condition gives a Jordan block, of the velocity matrix for $\mu\neq0$ and of the full dynamics for $\lambda\neq0$, and is an exceptional point of the corresponding larger family. Within the family of the model, only $\lambda=0$ is an exceptional point, where $\mu$ changes the Jordan blocks from $[2,2]$ to $[3,1]$. A change of the velocity couplings alone leaves the solutions unchanged exactly when both conditions hold.
\item \emph{The model: Lagrangians and Hamiltonian (Sections~\ref{sec:variational} and~\ref{sec:hamiltonian}).} For $\lambda\mu\neq0$ the pair of the model has no regular first-order Lagrangian. The substitution $x\to\Uop x$ gives a Lagrangian with second derivatives for $\lambda\neq0$, and a first-order Lagrangian with time-dependent coefficients for $\lambda=0$. In the Bateman doubling, $\Uop$ becomes a canonical transformation that removes $\mu$ from the Hamiltonian.
\end{itemize}
The rest of the paper is organized as follows. In Section~\ref{sec:model} we introduce the notation, solve~\eqref{eq:model} and show that it is similar to its $\mu=0$ version. In Section~\ref{sec:origin} we derive equations with reflection from local pairs of oscillators and give an electronic realization of the model. Section~\ref{sec:spectral} discusses the Jordan blocks of the pair and the exceptional points of the corresponding families. In Section~\ref{sec:variational} we study Lagrangians of the pair and of the equation, and in Section~\ref{sec:hamiltonian} we construct the Bateman Lagrangian and Hamiltonian and remove $\mu$ by a canonical transformation. In Section~\ref{sec:family} we ask when the solutions of the general linear equation determine its coefficients. We conclude with a few remarks and open problems in Section~\ref{sec:conclusion}. The pair with balanced gain and loss and some technical details are collected in the appendices.

%======================================================================
\section{The equation and its similarity to the \texorpdfstring{$\mu=0$}{mu=0} equation}\label{sec:model}

\subsection{The linear equation with reflection}\label{sec:notation}

We study the general linear equation with reflection
\begin{equation}
  \ddot x(t)=a\,x(t)+b\,x(-t)+c\,\dot x(t)+d\,\dot x(-t),
  \label{eq:general-linear}
\end{equation}
with real constants $a,b,c,d$. The model~\eqref{eq:model} is the case $a=\lambda$, $b=-\lambda$, $c=\mu$, $d=-\mu$, and its $\mu=0$ version is
\begin{equation}
  \ddot x(t)=\lambda\bigl[x(t)-x(-t)\bigr].
  \label{eq:mu0}
\end{equation}
All these equations are considered on the whole real line. A twice differentiable solution of~\eqref{eq:general-linear} is smooth, because the right-hand side is then differentiable, and so on by induction. We therefore work with smooth functions.

Let us introduce the notation. We write $D=\dd/\dd t$ and $(Rf)(t)=f(-t)$. Then
\begin{equation*}
  R^2=1,\qquad DR=-RD,\qquad Rt=-tR,
\end{equation*}
where $t$ denotes multiplication by $t$. In particular $\dot x(-t)=(RDx)(t)$, while $\dd[x(-t)]/\dd t=(DRx)(t)=-\dot x(-t)$. For a function $x$ we write $u=(1+R)x$ and $v=(1-R)x$, so that $u$ is even, $v$ is odd, and $x=\frac12(u+v)$. The three relations above give
\begin{gather*}
  (1+R)D=D(1-R),\qquad (1-R)D=D(1+R),\qquad (1+R)t=t(1-R),\\
  (1+R)(1-R)=0,\qquad (1\pm R)^2=2(1\pm R).
\end{gather*}
In words: $D$ and multiplication by $t$ map even functions to odd ones, and odd functions to even ones.

We use the pairing $\langle y,x\rangle=\int_{-T}^{T}y(t)\,x(t)\,\dd t$. The formal adjoint $A^\dagger$ of an operator $A$ is defined by $\langle y,Ax\rangle=\langle A^\dagger y,x\rangle$ for all functions that vanish, together with their derivatives, at $t=\pm T$. Integration by parts and the substitution $t\to-t$ give $D^\dagger=-D$, $R^\dagger=R$ and $t^\dagger=t$, and hence $(RD)^\dagger=RD$.

In operator form, equation~\eqref{eq:general-linear} reads $\E x=0$ with $\E=D^2-a-bR-cD-dRD$. Its adjoint is $\E^\dagger=D^2-a-bR+cD-dRD$, so
\begin{equation*}
  \E-\E^\dagger=-2cD:
\end{equation*}
the reflected velocity term $d\,\dot x(-t)$ is self-adjoint, and the local term $c\,\dot x(t)$ is not. For the model~\eqref{eq:model},
\begin{equation*}
  \E=D^2-\lambda(1-R)-\mu D(1+R),\qquad \E^\dagger=D^2-\lambda(1-R)+\mu D(1-R),
\end{equation*}
while its $\mu=0$ version~\eqref{eq:mu0} has the self-adjoint operator $\E_0=D^2-\lambda(1-R)$.

\subsection{Even and odd parts}\label{sec:evenodd}

Following~\cite{CabadaTojo2015,Ma2024,Mandal2025,Ma2024Pramana}, we split a solution of~\eqref{eq:model} into its even and odd parts $u=(1+R)x$ and $v=(1-R)x$. Applying $1+R$ and $1-R$ to $\E x=0$, and using $(1+R)D=D(1-R)$, $(1+R)(1-R)=0$ and $(1-R)^2=2(1-R)$, we obtain
\begin{equation}
  \ddot u=0,\qquad \ddot v=2\lambda v+2\mu\dot u.
  \label{eq:uv-model}
\end{equation}
Conversely, $x=\frac12(u+v)$ solves~\eqref{eq:model} if $u$ and $v$ solve~\eqref{eq:uv-model} with $u$ even and $v$ odd. Since $u$ is even, $\ddot u=0$ implies that $u=2C$ is constant, so that $\dot u=0$ and $\ddot v=2\lambda v$. Since $v$ is odd, the general solution of~\eqref{eq:model} is
\begin{equation}
  x(t)=\begin{cases}
  C+A\sinh\bigl(\sqrt{2\lambda}\,t\bigr), & \lambda>0,\\
  C+At, & \lambda=0,\\
  C+A\sin\bigl(\sqrt{-2\lambda}\,t\bigr), & \lambda<0,
  \end{cases}
  \label{eq:solution}
\end{equation}
with the constants fixed by $x(0)$ and $\dot x(0)$.\footnote{\label{fn:cauchy}For $\lambda<0$ and $\cos(\sqrt{-2\lambda}\,t_0)=0$, however, every solution has $\dot x(t_0)=0$. So the Cauchy problem at $t_0$ has either no solution or infinitely many, cf.~\cite{Mandal2025,Ma2024Pramana,Ma2025}.} The $\mu$-dependent part of $\E x$ is $\mu D(1+R)x=\mu\dot u$, which vanishes on solutions. Thus~\eqref{eq:model} has the same solutions as~\eqref{eq:mu0}.

\subsection{Similarity to the \texorpdfstring{$\mu=0$}{mu=0} equation}\label{sec:similarity}

The independence of $\mu$ can be stated as a similarity of operators. Define
\begin{equation}
  \Uop=1+\frac{\mu}{2\lambda}D(1+R)\quad(\lambda\neq0),\qquad
  \Uop=1-\frac{\mu}{2}\,t(1+R)\quad(\lambda=0),
  \label{eq:U}
\end{equation}
that is,
\begin{equation}
  (\Uop x)(t)=x(t)+\frac{\mu}{2\lambda}\bigl[\dot x(t)-\dot x(-t)\bigr]\quad(\lambda\neq0),\qquad
  (\Uop x)(t)=x(t)-\frac{\mu}{2}\,t\,\bigl[x(t)+x(-t)\bigr]\quad(\lambda=0).
  \label{eq:U-explicit}
\end{equation}
Then
\begin{equation}
  \E_0\,\Uop=\Uop\,\E .
  \label{eq:intertwine}
\end{equation}
For $\lambda\neq0$, the identities $(1+R)(1-R)=0$, $(1\pm R)D=D(1\mp R)$ and $(1-R)^2=2(1-R)$ show that both sides equal $\E+\frac{\mu}{2\lambda}D^3(1+R)$. For $\lambda=0$ we use in addition $D^2t=tD^2+2D$, and both sides equal $\E-\frac{\mu}{2}\,tD^2(1+R)$. By $(1+R)D=D(1-R)$, $(1+R)t=t(1-R)$ and $(1-R)(1+R)=0$, we have $\bigl(D(1+R)\bigr)^2=\bigl(t(1+R)\bigr)^2=0$. So $\Uop$ differs from the identity by a nilpotent operator, its inverse is obtained by changing the sign of $\mu$, and
\begin{equation}
  \E=\Uop^{-1}\,\E_0\,\Uop .
  \label{eq:similarity}
\end{equation}
This relation has three consequences.

\emph{First, equation~\eqref{eq:model} is equation~\eqref{eq:mu0} for a new unknown.} A function $x$ solves~\eqref{eq:model} if and only if $\tilde x=\Uop x$ solves~\eqref{eq:mu0}, so $\mu$ enters only through this change of variables. By~\eqref{eq:U-explicit}, it acts only through the even part $u$ of $x$: $\tilde x=x+\frac{\mu}{2\lambda}\dot u$ for $\lambda\neq0$, and $\tilde x=x-\frac{\mu}{2}tu$ for $\lambda=0$.

\emph{Second, the solutions do not depend on $\mu$.} The solutions of~\eqref{eq:model} are $\Uop^{-1}$ applied to those of~\eqref{eq:mu0}. For $\lambda\neq0$, $\Uop$ acts on the solutions of~\eqref{eq:mu0} as the identity, since $u$ is constant.\footnote{By contrast, for the damped oscillator the substitution $x=e^{\mu t/2}\tilde x$ gives $\ddot{\tilde x}=-(\omega^2-\mu^2/4)\tilde x$, which changes every solution.} For $\lambda=0$ it maps the solution $C+At$ to $C+(A-\mu C)t$: it maps the set of solutions onto itself, but not as the identity. In both cases the solution of~\eqref{eq:model} with given $x(0)$ and $\dot x(0)$ is the same for all $\mu$.

\emph{Third, the equation itself depends on $\mu$, and so do the solutions of its adjoint.} A similarity transformation does not preserve self-adjointness. Since $\E_0$ is self-adjoint,~\eqref{eq:similarity} gives $\E^\dagger=\Uop^\dagger\E_0(\Uop^\dagger)^{-1}$, with
\begin{equation}
  \Uop^\dagger=1-\frac{\mu}{2\lambda}D(1-R)\quad(\lambda\neq0),\qquad
  \Uop^\dagger=1-\frac{\mu}{2}\,t(1-R)\quad(\lambda=0).
  \label{eq:Udag}
\end{equation}
The solutions of $\E^\dagger y=0$ are $y=\Uop^\dagger y_0$, where $y_0$ solves~\eqref{eq:mu0}. Unlike $\Uop$, the operator $\Uop^\dagger$ acts through the odd part $(1-R)y_0$, which does not vanish in general, so the adjoint solutions depend on $\mu$.

%======================================================================
\section{Physical origin}\label{sec:origin}

The similarity $\Uop$ explains the independence of $\mu$ algebraically. To see its physical meaning, we now show that equations with reflection describe local systems of two oscillators.

\subsection{Exchange and time reversal}\label{sec:reduction}

Consider two oscillators with coordinates $x_1(t)$ and $x_2(t)$ and equations of motion
\begin{equation}
  \ddot x_1=F(x_1,x_2,\dot x_1,\dot x_2),\qquad \ddot x_2=F(x_2,x_1,-\dot x_2,-\dot x_1).
  \label{eq:pair}
\end{equation}
This system is invariant under the exchange $\Pex$: $x_1\leftrightarrow x_2$ combined with time reversal $\Trev$: $t\mapsto-t$,
\begin{equation}
  \sigma=\Pex\Trev:\qquad x_1(t)\to x_2(-t),\qquad x_2(t)\to x_1(-t),
  \label{eq:sigma}
\end{equation}
since $\tilde x_1(t)=x_2(-t)$ and $\tilde x_2(t)=x_1(-t)$ obey the same equations. Here parity exchanges the two oscillators, and time reversal changes the sign of the velocities. This is the $\PT$ operation used for the active $LRC$ dimers of~\cite{Schindler2012}. Conversely, an autonomous system $\ddot x_k=F_k(x_1,x_2,\dot x_1,\dot x_2)$ is invariant under $\sigma$ if and only if $F_2(x_1,x_2,\dot x_1,\dot x_2)=F_1(x_2,x_1,-\dot x_2,-\dot x_1)$, that is, if it has the form~\eqref{eq:pair} with $F=F_1$.

If the initial data are invariant under $\sigma$,
\begin{equation}
  x_2(0)=x_1(0),\qquad \dot x_2(0)=-\dot x_1(0),
  \label{eq:PTdata}
\end{equation}
then $\sigma$ maps the solution to a solution with the same initial data. For $F$ with a uniquely solvable Cauchy problem, therefore, $x_2(t)=x_1(-t)$ on the whole maximal interval of existence, which is then symmetric about $t=0$: the second oscillator is the time-reversed first one. Substituting $x_2(t)=x(-t)$ and $\dot x_2(t)=-\dot x(-t)$ into~\eqref{eq:pair}, we obtain for $x\equiv x_1$ the nonlocal equation
\begin{equation}
  \ddot x(t)=F\bigl(x(t),x(-t),\dot x(t),-\dot x(-t)\bigr).
  \label{eq:reduced}
\end{equation}
The second equation of~\eqref{eq:pair} is then the first one at $-t$. Conversely, if $x$ solves~\eqref{eq:reduced}, then $(x(t),x(-t))$ solves~\eqref{eq:pair} with the data~\eqref{eq:PTdata}. Moreover, every equation of the form $\ddot x(t)=h\bigl(x(t),x(-t),\dot x(t),\dot x(-t)\bigr)$ arises in this way, with $F(p,q,r,w)=h(p,q,r,-w)$. So the solutions of~\eqref{eq:reduced} are in one-to-one correspondence with the solutions of~\eqref{eq:pair} with the data~\eqref{eq:PTdata}, and~\eqref{eq:reduced} has a unique solution for given $x(0)$ and $\dot x(0)$. We call these $\sigma$-invariant solutions $\PT$-symmetric.

In~\cite{Bender2013PRA} parity carries an extra sign,
\begin{equation}
  \sigma'=\Pex'\Trev:\qquad x_1(t)\to -x_2(-t),\qquad x_2(t)\to -x_1(-t),
  \label{eq:sigmap}
\end{equation}
which differs from $\sigma$ by $\bm x\to-\bm x$. A system invariant under $\sigma'$ becomes invariant under $\sigma$ when $x_2$ is replaced by $-x_2$, and data invariant under $\sigma'$,
\begin{equation}
  x_2(0)=-x_1(0),\qquad \dot x_2(0)=\dot x_1(0),
  \label{eq:PTpdata}
\end{equation}
give $x_2(t)=-x_1(-t)$ and again a nonlocal equation for $x\equiv x_1$. The system~\eqref{eq:pair} is invariant under both $\sigma$ and $\sigma'$ if and only if $F$ is odd, in particular if it is linear. For a linear pair, $\sigma$ is linear on the space of solutions, $\sigma^2=1$ and $\sigma'=-\sigma$. Every solution is therefore the unique sum
\begin{equation*}
  \bm x=\tfrac12(\bm x+\sigma\bm x)+\tfrac12(\bm x-\sigma\bm x)
\end{equation*}
of a $\PT$-symmetric and a $\sigma'$-invariant solution, and the initial data split accordingly into~\eqref{eq:PTdata} and~\eqref{eq:PTpdata}. The two parts obey two different nonlocal equations. For the pair of Section~\ref{sec:model-dimer} they are~\eqref{eq:model} and $\ddot x(t)=\lambda[x(t)+x(-t)]+\mu[\dot x(t)+\dot x(-t)]$.

$\PT$ symmetry of a solution is a property of an individual trajectory, and such solutions exist for all parameters, whereas the unbroken $\PT$ phase of~\cite{Bender2013PRA} is a property of the spectrum (Appendix~\ref{sec:bender}). The nonlocality of~\eqref{eq:reduced} is only apparent: $x(-t)$ is simply the present state of the second oscillator. This is the analogue, for ordinary differential equations, of the reductions in~\cite{Yang2018} and especially of the reverse-time reductions $r(x,t)=\pm q(x,-t)$ of the AKNS system~\cite{AblowitzMusslimani2017}, whose two signs correspond to $\sigma$ and $\sigma'$. In the language of~\cite{Lou2018}, the pair with $\PT$-symmetric data is an Alice--Bob system.

\subsection{Linear pairs}\label{sec:linear}

For the linear equation~\eqref{eq:general-linear}, Section~\ref{sec:reduction} gives $F(p,q,r,w)=ap+bq+cr-dw$. The local system~\eqref{eq:pair} is then
\begin{equation}
  \ddot{\bm x}=K\bm x+\Gamma\dot{\bm x},\qquad
  K=\begin{pmatrix}a&b\\ b&a\end{pmatrix},\qquad
  \Gamma=\begin{pmatrix}c&-d\\ d&-c\end{pmatrix},\qquad
  \bm x=(x_1,x_2).
  \label{eq:dimer}
\end{equation}
Its solutions are the sums of $\PT$-symmetric solutions, whose first oscillator obeys~\eqref{eq:general-linear}, and of $\sigma'$-invariant ones.

The velocity-dependent forces deliver the power $\dot{\bm x}^{\mathsf T}\Gamma\dot{\bm x}=c\,(\dot x_1^2-\dot x_2^2)$. We write $\Gamma=c\,\sigma_z+d\,J$, with $\sigma_z=\mathrm{diag}(1,-1)$ and $J=\bigl(\begin{smallmatrix}0&-1\\1&0\end{smallmatrix}\bigr)$. The symmetric term $c\,\sigma_z$ pumps energy into one oscillator and removes it from the other at the same rate: it is balanced gain and loss. The antisymmetric term $d\,J$ does no work and rotates the velocity in the $(x_1,x_2)$ plane: it is a gyroscopic coupling, such as a Lorentz or Coriolis force or an ideal gyrator~\cite{Tellegen1956}. No other velocity forces are allowed. Indeed, since $\Pex$ acts as $\sigma_x$ and $\Trev$ reverses the velocities, the symmetry $\sigma$ requires $\sigma_xK\sigma_x=K$ and $\sigma_x\Gamma\sigma_x=-\Gamma$, and among $\one$, $\sigma_x$, $\sigma_z$ and $J$ only $\sigma_z$ and $J$ anticommute with $\sigma_x$. The springs need one remark. Since $a x_1+b x_2=(a+b)x_1-b(x_1-x_2)$, each oscillator has an on-site spring of stiffness $-(a+b)$, and the two oscillators are coupled by a spring of stiffness $b$. So the diagonal entry $a$ of $K$ contains the coupling spring as well, and $a+b=0$ means that there is no on-site spring. Table~\ref{tab:dictionary} gives the meaning of all coefficients.

\begin{table}[t]
\centering
\caption{Coefficients of~\eqref{eq:general-linear} and the corresponding forces in the local system~\eqref{eq:dimer} (unit masses).}
\label{tab:dictionary}
\begin{tabular}{@{}l l l l@{}}
\toprule
Term in~\eqref{eq:general-linear} & Force on $x_1$ & Force on $x_2$ & Physical element \\
\midrule
$a\,x(t)+b\,x(-t)$ & $a x_1+b x_2$ & $a x_2+b x_1$ & \begin{tabular}[t]{@{}l@{}}on-site springs (stiffness $-(a+b)$)\\ and a coupling spring (stiffness $b$)\end{tabular} \\
$c\,\dot x(t)$ & $c\dot x_1$ & $-c\dot x_2$ & balanced gain and loss \\
$d\,\dot x(-t)$ & $-d\dot x_2$ & $d\dot x_1$ & gyroscopic coupling \\
\bottomrule
\end{tabular}
\end{table}

In terms of $u=x_1+x_2$ and $v=x_1-x_2$, which on $\PT$-symmetric solutions coincide with the even and odd parts $(1\pm R)x$ of Section~\ref{sec:notation}, we have
\begin{equation}
  \ddot u=(a+b)\,u+(c+d)\,\dot v,\qquad \ddot v=(a-b)\,v+(c-d)\,\dot u.
  \label{eq:modes}
\end{equation}
Since the pair is linear with constant coefficients, it has solutions $e^{st}\bm\xi$, where the characteristic exponents $s$ are the roots of $\det(s^2-\Gamma s-K)=0$. These are also the exponents of~\eqref{eq:general-linear}. With $e^{st}\bm\xi$, the pair has the solution $e^{-st}\sigma_x\bm\xi$, its image under $\sigma$, so for $s\neq0$ it has the $\PT$-symmetric solution $e^{st}\bm\xi+e^{-st}\sigma_x\bm\xi$. Its first component, $x(t)=\xi_1e^{st}+\xi_2e^{-st}$, solves~\eqref{eq:general-linear}. Conversely, every solution of~\eqref{eq:general-linear} is the first component of a solution of the pair, so it is built from the exponents of the pair. When the exponents are $\pm s_1$ and $\pm s_2$ with $s_1$ and $s_2$ nonzero and $s_1^2\neq s_2^2$, the two pairs give two solutions of~\eqref{eq:general-linear}, and these span all its solutions. For $s=0$ the combination $\bm\xi+\sigma_x\bm\xi$ can vanish, and then, as at coinciding exponents, the $\PT$-symmetric solutions can contain polynomial factors in $t$. So the solutions of~\eqref{eq:general-linear} oscillate exactly when those of the pair do: the $\PT$ phases of the pair are those of~\eqref{eq:general-linear}. For $d=0$ the pair~\eqref{eq:dimer} is the $\PT$-symmetric pair of~\cite{Bender2013PRA}, which is realized with the active $LRC$ dimers of~\cite{Schindler2011,Schindler2012}; we discuss this example in Appendix~\ref{sec:bender}.

\subsection{The model}\label{sec:model-dimer}

Equation~\eqref{eq:model} corresponds to $a=\lambda$, $b=-\lambda$, $c=\mu$ and $d=-\mu$. The local system~\eqref{eq:dimer} is
\begin{equation}
  \ddot x_1=\lambda(x_1-x_2)+\mu(\dot x_1+\dot x_2),\qquad
  \ddot x_2=-\lambda(x_1-x_2)-\mu(\dot x_1+\dot x_2),
  \label{eq:local-model}
\end{equation}
with
\begin{equation}
  K=\lambda\begin{pmatrix}1&-1\\ -1&1\end{pmatrix},\qquad
  \Gamma=\mu\begin{pmatrix}1&1\\ -1&-1\end{pmatrix}=\mu\,\bm e\bm f^{\mathsf T},\qquad
  \bm e=\begin{pmatrix}1\\-1\end{pmatrix},\quad \bm f=\begin{pmatrix}1\\1\end{pmatrix}.
  \label{eq:KGamma}
\end{equation}
The matrix $K$ is a single spring between the two oscillators,\footnote{For $\lambda<0$ the spring has stiffness $-\lambda$; for $\lambda>0$ it is repulsive.} and $\Gamma=\mu(\sigma_z-J)$: balanced gain and loss and a gyroscopic coupling of the same strength. All forces are equal and opposite, so the total momentum $\pcm\equiv\dot x_1+\dot x_2$ is conserved. In the coordinates of~\eqref{eq:modes},
\begin{equation}
  \ddot u=0,\qquad \ddot v=2\lambda v+2\mu\dot u.
  \label{eq:model-modes}
\end{equation}
The centre of mass drives the relative coordinate, but not vice versa. Indeed, $\bm e$ is the only eigenvector of $\Gamma$, $\bm f^{\mathsf T}$ is its left eigenvector, and $\bm f^{\mathsf T}\bm e=0$. So the velocity-dependent force $\Gamma\dot{\bm x}=\mu\,\pcm\,\bm e$ depends only on the total momentum and acts only on the relative coordinate. Such a one-way coupling is the characteristic form of a Jordan block, and it links exceptional points with non-reciprocal interactions~\cite{Fruchart2021}. Exceptional points produced by unidirectional coupling have recently been realized in electronic circuits~\cite{Zhao2024}. Since the power $\mu(\dot x_1^2-\dot x_2^2)$ of these forces can be positive, a realization of the pair needs an active element (Section~\ref{sec:circuit}).

\begin{figure}[t]
\centering
\figfile{0.85\linewidth}{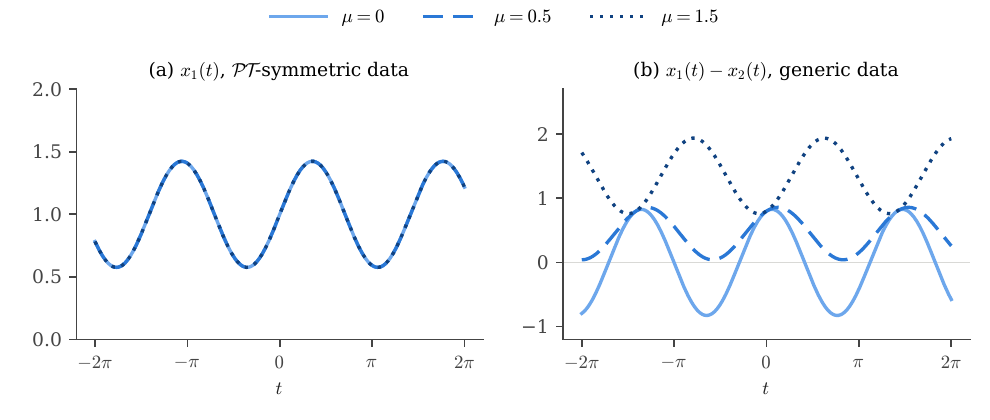}
\caption{Numerical solutions of~\eqref{eq:local-model} for $\lambda=-1$ and $\mu=0,\,0.5,\,1.5$. (a) $\PT$-symmetric data $x_1(0)=x_2(0)=1$, $\dot x_1(0)=-\dot x_2(0)=0.6$: the solutions coincide with~\eqref{eq:solution}. (b) Generic data $x_1(0)=1$, $x_2(0)=0.2$, $\dot x_1(0)=0.6$, $\dot x_2(0)=0.3$: the relative coordinate is shifted by $-\mu\pcm/\lambda=0,\,0.45,\,1.35$, with $\pcm=0.9$.}
\label{fig:dynamics}
\end{figure}

For $\PT$-symmetric data $\pcm(0)=0$, hence $\pcm=0$ at all times. The velocity-dependent forces then vanish, and~\eqref{eq:model-modes} reduces to $\ddot v=2\lambda v$ with $u=2x(0)$, which gives~\eqref{eq:solution}. The same holds for all data with $\pcm=0$. For $\pcm\neq0$ we have $\ddot v=2\lambda v+2\mu\pcm$. For $\lambda\neq0$ the relative coordinate is then shifted by $-\mu\pcm/\lambda$ (Figure~\ref{fig:dynamics}(b)), and for $\lambda=0$ it acquires the term $\mu\pcm t^2$. For $\lambda\neq0$ the shifted relative coordinate $\tilde v=v+(\mu/\lambda)\dot u$ obeys $\ddot{\tilde v}=2\lambda\tilde v$ for all initial data. This is the operator $\Uop$ of Section~\ref{sec:similarity} in the variables of the pair: by~\eqref{eq:U-explicit}, $\Uop$ leaves $u$ unchanged and replaces $v$ by $\tilde v$.

The total momentum is carried by the $\sigma'$-invariant solutions $x_2(t)=-x_1(-t)$, for which $\pcm=2\dot x_1(0)$ is in general nonzero. Their first oscillator obeys the second equation of Section~\ref{sec:reduction}, $\ddot x(t)=\lambda[x(t)+x(-t)]+\mu[\dot x(t)+\dot x(-t)]$, whose solutions do depend on $\mu$.

\subsection{Electronic realization}\label{sec:circuit}

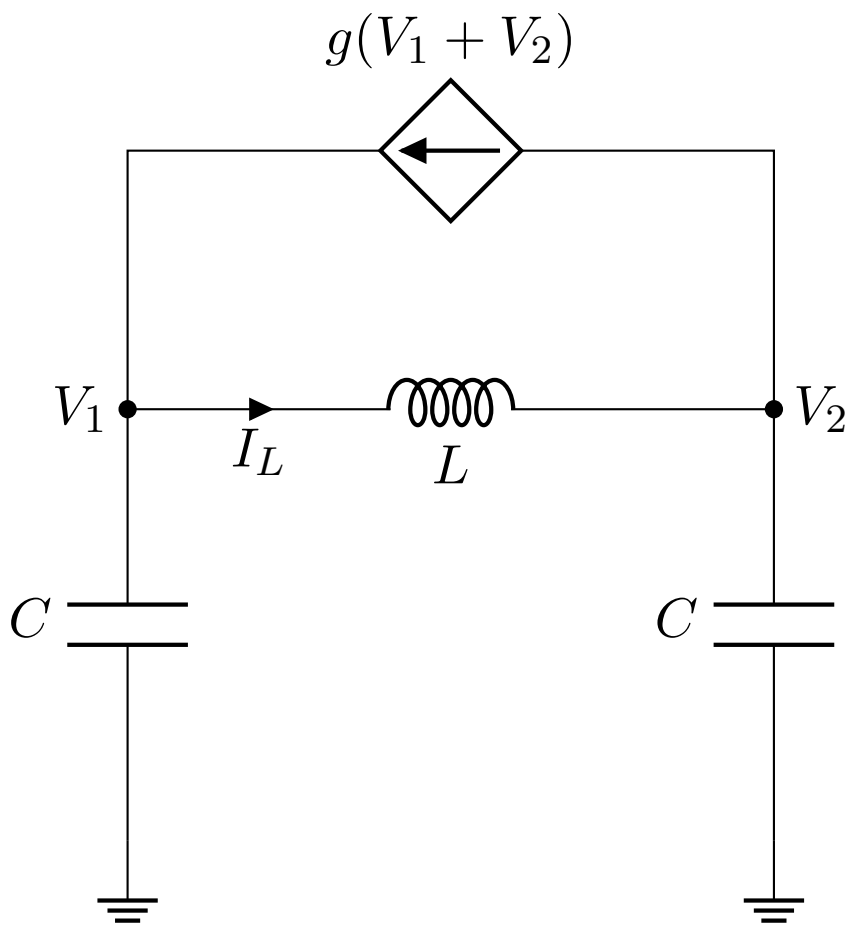
\begin{figure}[t]
\centering
\begin{circuitikz}[scale=0.95, transform shape, american currents]
  \draw (0,0) node[ground]{} to[C, l=$C$] (0,3) node[circ]{} node[left]{$V_1$};
  \draw (0,3) to[L, l_=$L$, i>_=$I_L$] (4.5,3) node[circ]{} node[right]{$V_2$};
  \draw (4.5,3) to[C, l_=$C$] (4.5,0) node[ground]{};
  \draw (4.5,3) -- (4.5,4.8) to[american controlled current source, l_=$g(V_1+V_2)$] (0,4.8) -- (0,3);
\end{circuitikz}
\caption{Electronic realization of~\eqref{eq:local-model}. Two grounded capacitors are coupled by an inductor; a transconductor connected between the two nodes injects $g(V_1+V_2)$ into node~1 and draws the same current from node~2.}
\label{fig:circuit}
\end{figure}

The pair~\eqref{eq:local-model} is realized by the circuit of Figure~\ref{fig:circuit}. Two nodes with voltages $V_1$ and $V_2$ are connected to ground through capacitors $C$ and to each other through an inductor $L$. A transconductor injects the current $g(V_1+V_2)$ into node~1 and draws the same current from node~2. With the flux linkages $\phi_k$, $\dot\phi_k=V_k$, as coordinates, Kirchhoff's current law gives
\begin{equation}
  C\ddot\phi_1=-\frac{\phi_1-\phi_2}{L}+g(\dot\phi_1+\dot\phi_2),\qquad
  C\ddot\phi_2=\frac{\phi_1-\phi_2}{L}-g(\dot\phi_1+\dot\phi_2),
  \label{eq:circuit}
\end{equation}
where the additive constants in $\phi_k$ are chosen so that the inductor current is $I_L=(\phi_1-\phi_2)/L$. This is~\eqref{eq:local-model} with $x_k=\phi_k$, $\lambda=-1/(LC)$ and $\mu=g/C$. The case $\lambda>0$ requires a negative inductance. The transconductor current splits into gain and loss, $gV_1$ into node~1 and $gV_2$ out of node~2~\cite{Schindler2011,Schindler2012}, and an ideal gyrator, $gV_2$ into node~1 and $gV_1$ out of node~2~\cite{Tellegen1956}. The two parts have the same conductance, which makes $\Gamma$ nilpotent.

$\PT$-symmetric data are $V_2(0)=-V_1(0)$ with no current in the inductor. The total charge $C(V_1+V_2)$ is conserved, so $V_1+V_2=0$ at all times: the transconductor carries no current, and the circuit is a passive $LC$ circuit that oscillates at $\sqrt{-2\lambda}$ for every $g$. The records of $\phi_1$ and $\phi_2$ for $t\ge0$ together give the solution $\phi_1(t)$ of~\eqref{eq:model} on the whole line, with $\phi_1(-t)=\phi_2(t)$. The effect of $g$ is seen only for generic data, through the shift $-\mu\pcm/\lambda$ of $\phi_1-\phi_2$ (Figure~\ref{fig:dynamics}(b)). The independence of $\mu$ requires $a+b=0$ and $c+d=0$ exactly, that is, no inductance to ground and equal conductances of the gain-and-loss and gyrator parts. Any deviation from these conditions makes the $\PT$-symmetric solutions depend on $g$ (Section~\ref{sec:family}). So for $\PT$-symmetric data the circuit tests the independence of $\mu$ only as a null result, and its accuracy is limited by the accuracy of these two conditions.

%======================================================================
\section{Jordan blocks and exceptional points}\label{sec:spectral}

The one-way coupling of Section~\ref{sec:model-dimer} is a Jordan block, and Jordan blocks appear at exceptional points. Within the family of linear pairs~\eqref{eq:dimer}, the model is singled out by two conditions, $c+d=0$ and $a+b=0$. We show that each of them is an exceptional-point condition. We stress that these are exceptional points of the larger families, not of the model family. In the model with $\lambda\neq0$ the characteristic exponents do not depend on $\mu$, and the Jordan block at $s=0$ is the free motion of the centre of mass, as in any system with a conserved total momentum.

\subsection{Exceptional points}\label{sec:EP}

A linear system $\dot z=Az$ has solutions $e^{st}\xi$ with $A\xi=s\xi$, and we call the eigenvalues $s$ of $A$ the characteristic exponents. If an exponent has fewer eigenvectors than its multiplicity, $A$ has a Jordan block of size larger than one there. A block of size $k$ gives solutions $e^{st}$ times polynomials in $t$ of degree up to $k-1$~\cite{Kato1995}. Following Kato~\cite{Kato1995}, an exceptional point of a family $A(\kappa)$ is a value $\kappa_0$ at which some eigenvalues coincide that are different for $\kappa\neq\kappa_0$. For several parameters, a point is exceptional if some eigenvalues coincide there that are different at generic nearby points of the family. Eigenvalues that coincide for all members of the family are a permanent degeneracy, not an exceptional point, so whether a point is exceptional depends on the family. In the physics literature the term usually also requires the eigenvectors to coalesce, so that $A(\kappa_0)$ is not diagonalizable~\cite{ElGanainy2018,Heiss2012}. We follow this usage and exclude coincidences at which $A(\kappa_0)$ is diagonalizable. At all remaining exceptional points in this paper the eigenvalues have a square-root branch point.

Two examples of Kato show the difference between a Jordan block and an exceptional point~\cite{Kato1995}. The equation $\ddot x=\kappa x$ has the exponents $\pm\sqrt{\kappa}$, which coalesce at $\kappa=0$ into a Jordan block of size two. So $\kappa=0$ is an exceptional point, and the free-particle solution $x_0+v_0t$ appears exactly there. By contrast, a free particle of mass $1/\kappa$, with $\dot x=\kappa p$ and $\dot p=0$, has a Jordan block for every $\kappa\neq0$. This family is permanently degenerate and has no exceptional point.

\subsection{The velocity coupling: \texorpdfstring{$c+d=0$}{c+d=0}}\label{sec:velocity}

Without springs, the velocities $\bm w=\dot{\bm x}$ of the pair~\eqref{eq:dimer} obey $\dot{\bm w}=\Gamma\bm w$, with $\Gamma=c\,\sigma_z+d\,J$. Since $\sigma_z$ and $J$ anticommute,
\begin{equation}
  \Gamma^2=(c^2-d^2)\,\one,
  \label{eq:Gamma2}
\end{equation}
so the eigenvalues of $\Gamma$ are $\pm\sqrt{c^2-d^2}$. They are imaginary for $c^2<d^2$, when the gyroscopic coupling dominates, and real for $c^2>d^2$, when gain and loss dominate. On the lines $c=\pm d$ they coincide, and~\eqref{eq:Gamma2} gives $\Gamma^2=0$. Gain and loss alone, or the gyroscopic coupling alone, is diagonalizable, but at equal strength and $c\neq0$ they combine into a single Jordan block of size two~\cite{Fruchart2021}, with the single eigenvector $(1,-1)$ for $c=-d$ and $(1,1)$ for $c=d$. Near these lines the eigenvalues have a square-root branch point. So in the two-parameter family $\Gamma(c,d)$ the points of the lines $c=\pm d$ with $c\neq0$ are exceptional points~\cite{Kato1995,Heiss2012,ElGanainy2018}. The point $c=d=0$, where $\Gamma=0$ is diagonalizable, is excluded.

The model has $\Gamma=\mu(\sigma_z-J)=\mu\,\bm e\bm f^{\mathsf T}$, which lies on the line $c=-d$. For $\mu\neq0$ it is therefore at an exceptional point of the family of velocity matrices. Within the model family, with $\mu$ as the only parameter, it is not: $\Gamma$ is nilpotent for every $\mu$, which is Kato's second example.

\subsection{The spring coupling: \texorpdfstring{$a+b=0$}{a+b=0}}\label{sec:springs}

The condition $a+b=0$ acts on the full dynamics. With $z=(\bm x,\dot{\bm x})$, the pair~\eqref{eq:dimer} becomes $\dot z=Az$ with $A=\bigl(\begin{smallmatrix}0&\one\\ K&\Gamma\end{smallmatrix}\bigr)$. Its characteristic exponents are the roots of $\det(s^2-\Gamma s-K)=0$, which is
\begin{equation}
  (s^2-a)^2-(c^2-d^2)\,s^2-b^2=0.
  \label{eq:charpoly}
\end{equation}
The two roots $s^2$ of this quadratic have the product $a^2-b^2$ and the sum $2a+c^2-d^2$. Let $b\neq0$ and $2a+c^2-d^2\neq0$. On the surface $a+b=0$ the product vanishes, so $s=0$ is a double exponent, and the other two exponents are $\pm\sqrt{2a+c^2-d^2}$. Since $K$ then has rank one, $A$ has rank three, and the exponent $0$ has a single eigenvector: it forms a Jordan block of size two. Off the surface the small root is $s^2\approx(a^2-b^2)/(2a+c^2-d^2)$, so the double exponent $0$ splits into a pair $\pm s$ with a square-root branch point. These points of the surface are therefore exceptional points of the full dynamics in the family $(a,b,c,d)$~\cite{Kato1995}.

For the model, $a=-b=\lambda$ and $c=-d=\mu$, so~\eqref{eq:charpoly} becomes $s^2(s^2-2\lambda)=0$, and both conditions above hold for $\lambda\neq0$. In the coordinates of~\eqref{eq:model-modes}, the eigenvector of the exponent $0$ describes both oscillators at rest at the same position. The generalized eigenvector gives the solution $u=t$, $v=-\mu/\lambda$: the centre of mass moves uniformly with $\pcm=1$, and the relative coordinate carries the shift $-\mu\pcm/\lambda$ of Section~\ref{sec:model-dimer}. So this Jordan block comes from the conservation of the total momentum. The parameter $\mu$ enters only the generalized eigenvector, so the Jordan form does not depend on $\mu$, in agreement with the time-independent $\Uop$ of Section~\ref{sec:similarity}. Within the model family the block is permanent. Near the model, however, the small root is $s^2\approx a+b$. So any on-site spring, for example an inductance to ground in the circuit of Section~\ref{sec:circuit}, splits the block, and every point of the model with $\lambda\neq0$ is an exceptional point of the full dynamics in the family $(a,b,c,d)$.

\subsection{The point \texorpdfstring{$\lambda=0$}{lambda=0}}\label{sec:jordan-model}

Within the model family itself there is one exceptional point. As $\lambda\to0$, the exponents $\pm\sqrt{2\lambda}$ coalesce with $0$, with a square-root branch point. So $\lambda=0$ is an exceptional point of the model family, for every $\mu$, and all four exponents vanish. Here $\mu$ matters. For $\mu=0$ the two oscillators are free: the exponent $0$ has two Jordan blocks of size two, and all solutions are at most linear in $t$. For $\mu\neq0$ there are still two eigenvectors, but the blocks have sizes three and one, and the solution $u=\pcm t$, $v=\mu\pcm t^2$ of Section~\ref{sec:model-dimer} is quadratic in $t$. $\PT$-symmetric solutions have $\pcm=0$ and stay linear in $t$. Because $\mu$ changes the Jordan blocks at $\lambda=0$, no time-independent linear change of variables can remove it there. This is why $\Uop=1-\frac{\mu}{2}t(1+R)$ depends on time.

%======================================================================
\section{Lagrangian formulation}\label{sec:variational}

We now ask whether the linear equation with reflection, and in particular~\eqref{eq:model}, has a Lagrangian. We first recall what this means for ordinary differential equations (Section~\ref{sec:lagrangians}). Since an equation with reflection is the $\PT$-symmetric part of a local pair (Section~\ref{sec:origin}), we then translate nonlocal Lagrangians into Lagrangians of the pair (Section~\ref{sec:nonlocal-L}).

\subsection{Lagrangians and the inverse problem}\label{sec:lagrangians}

We call $L$ a Lagrangian for a system if the stationary points of its action $\int L\,\dd t$, with variations vanishing at the end points, are exactly the solutions of the system. A Lagrangian $L(t,\bm x,\dot{\bm x})$ is regular if its velocity Hessian $W_{ij}=\partial^2L/\partial\dot x_i\partial\dot x_j$ is nonsingular~\cite{Douglas1941,Santilli1978}. Its Euler--Lagrange expressions $E_i(L)=\partial L/\partial x_i-(\dd/\dd t)\,\partial L/\partial\dot x_i$ are then linear in the accelerations, with coefficient matrix $-W$. For a system
\begin{equation}
  \ddot{\bm x}=\bm f(t,\bm x,\dot{\bm x}),
  \label{eq:newton}
\end{equation}
the inverse problem of the calculus of variations asks whether there is a regular $L(t,\bm x,\dot{\bm x})$ such that
\begin{equation}
  E(L)=-W\bigl(\ddot{\bm x}-\bm f\bigr)
  \label{eq:inverse}
\end{equation}
holds identically in $(t,\bm x,\dot{\bm x},\ddot{\bm x})$~\cite{Douglas1941,Sarlet1982}. The velocity Hessian $W$ then acts as a multiplier that makes the system variational, and since $W$ is nonsingular, the stationary points of the action are exactly the solutions of~\eqref{eq:newton}, for all initial data. We then call $L$ a regular first-order Lagrangian for~\eqref{eq:newton}. A nonsingular matrix $W(t,\bm x,\dot{\bm x})$ is such a multiplier if and only if it satisfies the Helmholtz conditions~\cite{Douglas1941,Sarlet1982}; see~\cite{Prince2026} for recent developments, including systems with constraints. Section~\ref{sec:EP-theorem} states them for the linear systems we need, and Section~\ref{sec:nonlocal-L} extends the definition to equations with reflection.

If $L$ is quadratic, its Euler--Lagrange equations are linear, and their operator is formally self-adjoint, as follows from Jacobi's theorem on the second variation~\cite{Douglas1941,Santilli1978}. So a linear equation $\E x=0$ with $\E\neq\E^\dagger$ is not itself the Euler--Lagrange equation of a Lagrangian depending on $x$ alone. It can still have a regular first-order Lagrangian after multiplication by a multiplier $W$.

\subsection{Nonlocal and folded Lagrangians}\label{sec:nonlocal-L}

Let $L_1$ depend on $x(t)$, $x(-t)$, $\dot x(t)$ and $\dot x(-t)$, and consider the action $S=\int_{-T}^{T}L_1\,\dd t$, with variations that vanish at $t=\pm T$. Changing variables $t\to-t$ in the terms with reflected arguments, we obtain the Euler--Lagrange equation\footnote{In the last term of~\eqref{eq:nonlocal-EL} the substitution $t\to-t$ is made before the differentiation.}
\begin{equation}
  \frac{\partial L_1}{\partial x(t)}-\frac{\dd}{\dd t}\frac{\partial L_1}{\partial\dot x(t)}
  +\Bigl[\frac{\partial L_1}{\partial x(-t)}\Bigr]_{t\to-t}
  -\frac{\dd}{\dd t}\Bigl[\frac{\partial L_1}{\partial\dot x(-t)}\Bigr]_{t\to-t}=0.
  \label{eq:nonlocal-EL}
\end{equation}
This is the time-reflection analogue of the generalized Euler--Lagrange equations for Lagrangian densities that depend on a field at $x$ and $-x$~\cite{Pazarci2026}. These extend earlier variational formulations of nonlocal equations: for a shift $x\to x+\Delta x$~\cite{FujiokaEspinosa2019}, and for a reflection without derivatives at $-x$~\cite{VelascoJuanFujioka2022}.

Equivalently, we split the integral at $t=0$ and set $x_1(t)=x(t)$, $x_2(t)=x(-t)$ for $t\in[0,T]$. Then $S=\int_0^TL\,\dd t$ with the local Lagrangian
\begin{equation}
  L=L_1(x_1,x_2,\dot x_1,-\dot x_2)+L_1(x_2,x_1,-\dot x_2,\dot x_1),
  \label{eq:folded}
\end{equation}
which is invariant under $\sigma$. Its Euler--Lagrange equations with respect to $x_1$ and $x_2$ are~\eqref{eq:nonlocal-EL} at $t$ and at $-t$, and the continuity of $x$ and $\dot x$ at $t=0$ gives the $\PT$-symmetric data~\eqref{eq:PTdata}. If the common value $x_1(0)=x_2(0)$ is not fixed, the variation also gives the boundary term $-(\partial L/\partial\dot x_1+\partial L/\partial\dot x_2)\,\delta x_1(0)$. Its vanishing is a natural boundary condition, which for the Lagrangian~\eqref{eq:Lx} below is the second condition in~\eqref{eq:PTdata}. Thus a nonlocal Lagrangian $L_1$ corresponds to a $\sigma$-invariant Lagrangian for the local system. We call $L_1$ a regular first-order Lagrangian for an equation with reflection if its folded form~\eqref{eq:folded} is one, in the sense of Section~\ref{sec:lagrangians}, for the corresponding local system.

The reflection is a special kind of nonlocality. For general Lagrangians that are nonlocal in time, the Euler--Lagrange equations are functional-differential equations, and in general no finite set of initial data determines a solution~\cite{LlosaVives1994}; see~\cite{HerediaLlosa2025} for a recent Hamiltonian treatment. For a reflection, folding gives a local system, and by Section~\ref{sec:reduction} the solutions are determined by $x(0)$ and $\dot x(0)$.

\label{sec:L-mu0}As an example, consider the model. By Section~\ref{sec:model-dimer}, the $\PT$-symmetric solutions of the pair~\eqref{eq:local-model} coincide with those of its $\mu=0$ version, whose Lagrangian is $\frac12(\dot x_1^2+\dot x_2^2)+\frac{\lambda}{2}(x_1-x_2)^2$. Returning to the original variable, we obtain the nonlocal Lagrangian
\begin{equation}
  L_1=\tfrac12\bigl(\dot x(t)^2+\dot x(-t)^2\bigr)+\tfrac{\lambda}{2}\bigl(x(t)-x(-t)\bigr)^2 .
  \label{eq:Lx}
\end{equation}
Its folded form~\eqref{eq:folded} is twice the Lagrangian of the pair, and its Euler--Lagrange equation~\eqref{eq:nonlocal-EL} is $-2\E_0x=0$, that is,~\eqref{eq:mu0}. On $\PT$-symmetric solutions the energy of the $\mu=0$ pair is $\dot x(0)^2$.

So $L_1$ describes the solutions of~\eqref{eq:model}, but its Euler--Lagrange equation is~\eqref{eq:mu0}, and its folded form is the pair~\eqref{eq:local-model} with $\mu=0$, whereas for $\mu\neq0$ the pair has solutions that depend on $\mu$ (Figure~\ref{fig:dynamics}(b)). By the definitions above,~\eqref{eq:Lx} is therefore a regular first-order Lagrangian for~\eqref{eq:mu0}, but not for~\eqref{eq:model}. In the rest of this section we ask whether~\eqref{eq:model} has one.

\subsection{Autonomous quadratic Lagrangians}\label{sec:autonomous}

We first look for a Lagrangian of the pair~\eqref{eq:dimer} of the form
\begin{equation}
  L=\tfrac12\,\dot{\bm x}^{\mathsf T}W\dot{\bm x}+\dot{\bm x}^{\mathsf T}Q\bm x-\tfrac12\,\bm x^{\mathsf T}V\bm x,\qquad W=W^{\mathsf T},\quad V=V^{\mathsf T},\quad\det W\neq0.
  \label{eq:quadL}
\end{equation}
Its Euler--Lagrange equations, $W\ddot{\bm x}+(Q-Q^{\mathsf T})\dot{\bm x}+V\bm x=0$, coincide with $W(\ddot{\bm x}-\Gamma\dot{\bm x}-K\bm x)=0$ if $W\Gamma$ is antisymmetric and $WK$ is symmetric. For $W=\bigl(\begin{smallmatrix}p&q\\ q&r\end{smallmatrix}\bigr)$, the first condition gives
\begin{equation}
  pc+qd=0,\qquad qd+rc=0,\qquad d(r-p)=0.
  \label{eq:WGamma}
\end{equation}
For $d\neq0$ these conditions give $r=p$ and $q=-pc/d$, so that $\det W=p^2(d^2-c^2)/d^2$. For $d=0$ and $c\neq0$ they give $W=q\,\sigma_x$, and for $c=d=0$ we can take $W=\one$. In all cases $W$ has, or can be chosen to have, equal diagonal entries. Then $W$ commutes with $K=a\one+b\,\sigma_x$, so $WK$ is symmetric as well. Therefore the pair~\eqref{eq:dimer} has an autonomous quadratic Lagrangian unless $c=\pm d\neq0$, that is, unless $\Gamma$ is at an exceptional point of the family $\Gamma(c,d)$ (Section~\ref{sec:velocity}). In particular, the model has no such Lagrangian for $\mu\neq0$. For balanced gain and loss alone ($d=0$), the kinetic term is proportional to $\dot x_1\dot x_2$, as in Bateman's Lagrangian for the damped oscillator and its time-reversed partner~\cite{Bateman1931} and in the Lagrangian obtained from the Hamiltonian of the pair of~\cite{Bender2013PRA}.

For all $c^2\neq d^2$ we can take the single choice $W=\bigl(\begin{smallmatrix}d&-c\\ -c&d\end{smallmatrix}\bigr)$. In terms of $x$, it gives the nonlocal Lagrangian
\begin{equation}
  L_1=\frac d2\,\dot x(t)^2+\frac c2\,\dot x(t)\dot x(-t)+\frac{d^2-c^2}{2}\,x(t)\dot x(-t)+\frac{ad-bc}{2}\,x(t)^2+\frac{bd-ac}{2}\,x(t)x(-t),
  \label{eq:L1-general}
\end{equation}
whose Euler--Lagrange equation~\eqref{eq:nonlocal-EL} is $-(d-cR)\E x=0$, with $\E$ the operator of~\eqref{eq:general-linear}. The operator $d-cR$ is the multiplier $W$ written for $x$: on $\PT$-symmetric solutions the second equation of the pair is the first one at $-t$, so the first row of $W(\ddot{\bm x}-\Gamma\dot{\bm x}-K\bm x)$ is $d(\E x)(t)-c(\E x)(-t)$.

The Euler--Lagrange equation is equivalent to~\eqref{eq:general-linear} only if $d-cR$ is invertible. Since $(d-cR)(d+cR)=d^2-c^2$, this holds for $c^2\neq d^2$. At $c=\pm d$, however, the multiplier becomes $d(1\mp R)$, which annihilates all even or all odd functions. For the model, $d-cR=-\mu(1+R)$. By the identities of Section~\ref{sec:notation}, $(1+R)\E=D^2(1+R)$, so the Euler--Lagrange equation is $\ddot u=0$ alone, and the equation for $v$ is lost. This is the exceptional point of $\Gamma$ seen from the reflection equation.

\subsection{Lagrangians of arbitrary form}\label{sec:EP-theorem}

We now allow any regular first-order Lagrangian $L(t,x_1,x_2,\dot x_1,\dot x_2)$. For two degrees of freedom the inverse problem was solved in general by Douglas~\cite{Douglas1941}, and we compare with his classification below. For the linear pair the conditions can also be solved directly. By~\eqref{eq:inverse}, if $L$ is a Lagrangian for the pair~\eqref{eq:dimer}, its Euler--Lagrange equations are $-W(\ddot{\bm x}-\Gamma\dot{\bm x}-K\bm x)=0$, where the multiplier $W$ is the velocity Hessian. In Section~\ref{sec:autonomous} it was constant, while for the damped oscillator the time-dependent multiplier $e^{-\mu t}$ gives a Lagrangian~\cite{Bateman1931,Kanai1948}. For the pair, the Helmholtz conditions take a simple form, because $\bm f=K\bm x+\Gamma\dot{\bm x}$ is linear with constant coefficients. A nonsingular matrix $W(t,\bm x,\dot{\bm x})$ is a multiplier if and only if $W$ is symmetric, $\partial W_{ij}/\partial\dot x_k$ is symmetric in all three indices, $DW/Dt=-\frac12(W\Gamma+\Gamma^{\mathsf T}W)$, and $W\Phi$ is symmetric, where $\Phi=-K-\frac14\Gamma^2$ for this system~\cite{Douglas1941,Sarlet1982}. Here $D/Dt=\partial/\partial t+\dot x_k\,\partial/\partial x_k+(K\bm x+\Gamma\dot{\bm x})_k\,\partial/\partial\dot x_k$ is the derivative along solutions of~\eqref{eq:dimer}. By~\eqref{eq:Gamma2}, $\Gamma^2$ is a multiple of the identity, so $W\Phi$ is symmetric exactly when $WK$ is, and since $W$ and $K$ are symmetric, this means that they commute. We use the conditions
\begin{equation}
  W=W^{\mathsf T},\qquad \frac{DW}{Dt}=-\frac12\bigl(W\Gamma+\Gamma^{\mathsf T}W\bigr),\qquad [W,K]=0,
  \label{eq:helmholtz-pair}
\end{equation}
which must hold at every point $(t,\bm x,\dot{\bm x})$.

For $W=\bigl(\begin{smallmatrix}M_1&N\\ N&M_2\end{smallmatrix}\bigr)$, the condition $[W,K]=0$ reads $b(M_1-M_2)=0$, and the differential equation in~\eqref{eq:helmholtz-pair} reads
\begin{equation}
  \frac{DM_1}{Dt}=-cM_1-dN,\qquad \frac{DM_2}{Dt}=cM_2+dN,\qquad \frac{DN}{Dt}=\frac d2\,(M_1-M_2).
  \label{eq:Mdot}
\end{equation}
For $b\neq0$ we have $M_1=M_2$ at every point, hence $DM_1/Dt=DM_2/Dt$, and the first two equations give $cM_1+dN=0$. At $c=\pm d\neq0$ this means $N=\mp M_1$, so $\det W=M_1^2-N^2=0$ and the multiplier is singular. For $b=0$, $\det W$ is conserved because $\operatorname{tr}\Gamma=0$, and the solution of~\eqref{eq:Mdot} with $W(0)=\one$ is an invertible multiplier that depends only on $t$. Together with Section~\ref{sec:autonomous}, this gives the following result. \emph{The pair~\eqref{eq:dimer} has a regular first-order Lagrangian unless $b\neq0$ and $c=\pm d\neq0$. In this excluded case it has no such Lagrangian at all: quadratic or not, with constant or time-dependent coefficients.}

This result agrees with Douglas's classification~\cite{Douglas1941}, whose assumption of analyticity holds here. For the pair~\eqref{eq:dimer}, the rows of Douglas's matrix~\cite[eqs.~(3.1)--(3.4)]{Douglas1941} are $(-2b,0,2b)$, $(2bc,4bd,2bc)$ and $\bigl(2b(d^2-c^2),0,2b(c^2-d^2)\bigr)$. For $b=0$ the matrix vanishes, which is his Case~I, where a Lagrangian always exists. For $b\neq0$ and $c=d=0$ it has rank one, which is his Case~IIa. For $b\neq0$ and $(c,d)\neq(0,0)$ it has rank two, which is his Case~III, and his invariant $D=\Delta_1\Delta_3-\Delta_2^2$ equals $64b^4(d^2-c^2)$. So the excluded case $b\neq0$, $c=\pm d\neq0$ is exactly his Case~IIIb, in which no Lagrangian exists~\cite[Theorem~VII]{Douglas1941}. For $b\neq0$ and $c^2\neq d^2$ the pair is in Case~IIIa, where a Lagrangian, if it exists, is unique up to a constant factor and a total derivative~\cite[Theorem~VI]{Douglas1941}. It is therefore the autonomous Lagrangian of Section~\ref{sec:autonomous}. The direct derivation above shows the mechanism in terms of the pair: the spring coupling forces the two diagonal entries of the multiplier to be equal, and at the exceptional point of $\Gamma$ the Helmholtz conditions then make the multiplier singular.

For the model, $b=-\lambda$ and $c=-d=\mu$, so for $\lambda\mu\neq0$ the model belongs to the excluded class. This is a statement about the pair~\eqref{eq:local-model}. It applies to~\eqref{eq:model} only through the definition of Section~\ref{sec:nonlocal-L}, since the solutions of~\eqref{eq:model} alone are the stationary points of~\eqref{eq:Lx} (Section~\ref{sec:L-mu0}). Note that the pair is fixed by the form of the equation, not by its solutions. Two equations with the same solutions can have different pairs, and one of the pairs can have a Lagrangian while the other has none (Section~\ref{sec:family}).

In terms of $x$, a multiplier $W(t)$ of the pair that is compatible with $\sigma$ corresponds to the multiplier $M(t)+N(t)R$ of~\eqref{eq:general-linear}, with $M_1(t)=M(t)$, $M_2(t)=M(-t)$ and $N$ even, because on $\PT$-symmetric solutions the second row of the pair is the first row at $-t$. The reflection term $N(t)R$ is needed even for $\lambda=0$. With $N=0$,~\eqref{eq:Mdot} requires $M=e^{-\mu t}$ because of the local term $\mu\dot x(t)$, and an even $M$ because of the reflected term $\mu\dot x(-t)$.

For $\lambda=0$ the condition $[W,K]=0$ is empty, and~\eqref{eq:helmholtz-pair} gives $W(t)=e^{-\Gamma^{\mathsf T}t/2}\,W(0)\,e^{-\Gamma t/2}$. Since $\Gamma^2=0$, this multiplier is a polynomial rather than an exponential. For $W(0)=\one$ we obtain $M=1-\mu t+\mu^2t^2/2$, $N=\mu^2t^2/2$ and $\det W=1$. This multiplier has a simple origin. For $\lambda=0$ the relative coordinate of the pair is uniformly accelerated (Section~\ref{sec:jordan-model}), and in the accelerated frame $\tilde v=v-\mu tu$ the pair moves freely. The free Lagrangian $\frac14\dot u^2+\frac14\dot{\tilde v}^2$, written in the original variables, is
\begin{equation}
  L=\tfrac14\,\dot u^2+\tfrac14\bigl(\dot v-\mu u-\mu t\dot u\bigr)^2,\qquad u=x_1+x_2,\quad v=x_1-x_2.
  \label{eq:Lacc}
\end{equation}
Its velocity Hessian is exactly the $W(t)$ found above, and its Euler--Lagrange equations are the equations~\eqref{eq:local-model} with $\lambda=0$, multiplied by $-W(t)$.

\subsection{A Lagrangian with second derivatives}\label{sec:second-order}

The result of Section~\ref{sec:EP-theorem} concerns regular first-order Lagrangians. If the Lagrangian may also depend on the accelerations, the similarity~\eqref{eq:similarity} gives a Lagrangian for $\lambda\neq0$. Since $x$ solves~\eqref{eq:model} if and only if $\tilde x=\Uop x$ solves~\eqref{eq:mu0}, we substitute
\begin{equation}
  \tilde x(t)=x(t)+\beta\,\dot u(t),\qquad \beta=\frac{\mu}{2\lambda},
  \label{eq:xtilde}
\end{equation}
into the Lagrangian~\eqref{eq:Lx} of~\eqref{eq:mu0}. With $u=x(t)+x(-t)$ and $v=x(t)-x(-t)$, we obtain
\begin{equation}
  L_1=\tfrac12\bigl(\dot x(t)+\beta\ddot u\bigr)^2+\tfrac12\bigl(\dot x(-t)+\beta\ddot u\bigr)^2+\tfrac{\lambda}{2}\bigl(v+2\beta\dot u\bigr)^2 ,
  \label{eq:L2}
\end{equation}
a quadratic Lagrangian with constant coefficients that depends on $x(\pm t)$, $\dot x(\pm t)$ and $\ddot x(\pm t)$. Lagrangians with second derivatives and their variational symmetries are discussed, for example, in~\cite{Coban2023}.

Its Euler--Lagrange equation is $\Uop^\dagger\Uop\,\E x=0$. Although it contains the fourth derivative of $u$, the operator $\Uop^\dagger\Uop$ has an inverse that is again a differential operator, so the stationary points of~\eqref{eq:L2} are exactly the solutions of~\eqref{eq:model}, with the same initial data. The same holds for the folded form and the pair~\eqref{eq:local-model}, for all initial data. Still,~\eqref{eq:L2} is degenerate: it depends on the accelerations only through $\ddot u$, and its Hamiltonian formulation has a primary constraint. The details are given in Appendix~\ref{app:second-order}.

The construction does not contradict Section~\ref{sec:EP-theorem}, since the Euler--Lagrange equation reduces to~\eqref{eq:model} only after the factor $\Uop^\dagger\Uop$ is removed. For $\lambda=0$ the same construction, with $\Uop=1-\frac{\mu}{2}t(1+R)$, gives back the first-order Lagrangian~\eqref{eq:Lacc}, up to a factor of two, because this $\Uop$ contains no derivatives. Indeed, for $\lambda=0$ the factor $\Uop^\dagger\Uop=1-\mu t+\frac12\mu^2t^2+\frac12\mu^2t^2R$ is exactly the multiplier $M(t)+N(t)R$ of Section~\ref{sec:EP-theorem}, with the functions $M$ and $N$ found there from the Helmholtz conditions. For $\lambda\neq0$ it contains derivatives, so it is not a multiplier in the sense of Section~\ref{sec:EP-theorem}.

%======================================================================
\section{Bateman doubling and Hamiltonian}\label{sec:hamiltonian}

\subsection{The doubled Lagrangian}\label{sec:bateman}

The Lagrangians of Sections~\ref{sec:EP-theorem} and~\ref{sec:second-order} depend on $x$ alone, but for $\lambda=0$ the coefficients depend on time, and for $\lambda\neq0$ the Lagrangian depends on the accelerations and is degenerate. We now follow the standard route for non-self-adjoint linear equations~\cite{Bateman1931,FeshbachTikochinsky1977} and introduce an auxiliary coordinate $y$ with the action $-\int y\,\E x\,\dd t$. For all $\lambda$ and $\mu$, this gives a regular first-order Lagrangian with constant coefficients and a conserved Hamiltonian, the usual starting point for quantization~\cite{FeshbachTikochinsky1977,Celeghini1992,BlasoneJizba2004,Valdez2025,Beau2026}; see also~\cite{Bender2013PRA}. The price is the auxiliary coordinate, which obeys the adjoint equation and depends on $\mu$.

This doubling should not be confused with the pair of Section~\ref{sec:origin}. For $b=d=0$ the pair~\eqref{eq:dimer} is itself Bateman's dual system, whose second oscillator is the time-reversed image of the first~\cite{Bateman1931,FeshbachTikochinsky1977,BlasoneJizba2004}. Here the auxiliary coordinate $y$ doubles the pair once more, so the folded system has four coordinates.

We take, in the folded variables $x_1(t)=x(t)$, $x_2(t)=x(-t)$, $y_1(t)=y(t)$ and $y_2(t)=y(-t)$,
\begin{equation}
  L_1=\dot x_1\dot y_1-\frac{\mu}{2}\bigl(x_1\dot y_1-\dot x_1y_1\bigr)+\lambda x_1y_1
  -\frac{\lambda}{2}\bigl(x_1y_2+y_1x_2\bigr)+\frac{\mu}{2}\bigl(y_1\dot x_2+x_1\dot y_2\bigr).
  \label{eq:L1-doubled}
\end{equation}
Here the dots on $x_2$ and $y_2$ are derivatives of $x(-t)$ and $y(-t)$, so $\dot x_2=-\dot x(-t)$. Up to a total derivative and a term that is odd in $t$, $L_1=-y\,\E x$ (Appendix~\ref{app:bateman}), so $S=-\int_{-T}^{T}y\,\E x\,\dd t$ up to boundary terms. The folded Lagrangian~\eqref{eq:folded} is
\begin{equation}
  L=\dot x_1\dot y_1+\dot x_2\dot y_2+\lambda(x_1-x_2)(y_1-y_2)
  -\frac{\mu}{2}(x_1+x_2)(\dot y_1-\dot y_2)+\frac{\mu}{2}(y_1-y_2)(\dot x_1+\dot x_2).
  \label{eq:L-doubled}
\end{equation}
The term $\dot x_1\dot y_1+\dot x_2\dot y_2$ plays the role of the kinetic term, as in the Lagrangians of~\cite{FeshbachTikochinsky1977,Celeghini1992,BlasoneJizba2004}. Bateman's original Lagrangian has the opposite overall sign, which amounts to $y\to-y$~\cite{Bateman1931}.

Varying $y_1$ and $y_2$ gives the equations of motion~\eqref{eq:local-model} for $x$. Varying $x_1$ and $x_2$ gives the equations for $y$, the adjoint system
\begin{equation}
  \ddot y_1=\lambda(y_1-y_2)-\mu(\dot y_1-\dot y_2),\qquad
  \ddot y_2=-\lambda(y_1-y_2)-\mu(\dot y_1-\dot y_2),
  \label{eq:adjoint-system}
\end{equation}
that is, $\ddot{\bm y}=K\bm y-\Gamma^{\mathsf T}\dot{\bm y}$. This is the folded form of $\E^\dagger y=0$. Its velocity matrix $-\Gamma^{\mathsf T}$ has $c=d=-\mu$, so it lies on the other exceptional line $c=d$ of the family $\Gamma(c,d)$ (Section~\ref{sec:velocity}). Since $\Gamma^{\mathsf T}=\mu\,\bm f\bm e^{\mathsf T}$, the roles of the two modes are exchanged:
\begin{equation*}
  \ddot v_y=2\lambda v_y,\qquad \ddot u_y=-2\mu\dot v_y,
\end{equation*}
with $u_y=y_1+y_2$ and $v_y=y_1-y_2$. So the relative coordinate now drives the centre of mass.

\subsection{The auxiliary coordinate depends on \texorpdfstring{$\mu$}{mu}}\label{sec:aux}

For $\PT$-symmetric solutions of the doubled system, $y_2(t)=y(-t)$, so $u_y$ is even and $v_y$ is odd. The mode equation $\ddot u_y=-2\mu\dot v_y$ integrates to $\dot u_y+2\mu v_y=\mathrm{const}$. The left-hand side is odd, so the constant vanishes, and
\begin{equation*}
  \dot u_y=-2\mu v_y .
\end{equation*}
This does not vanish unless $\mu=0$ or $y$ is constant, so $y$ depends on $\mu$.

\begin{figure}[t]
\centering
\figfile{0.85\linewidth}{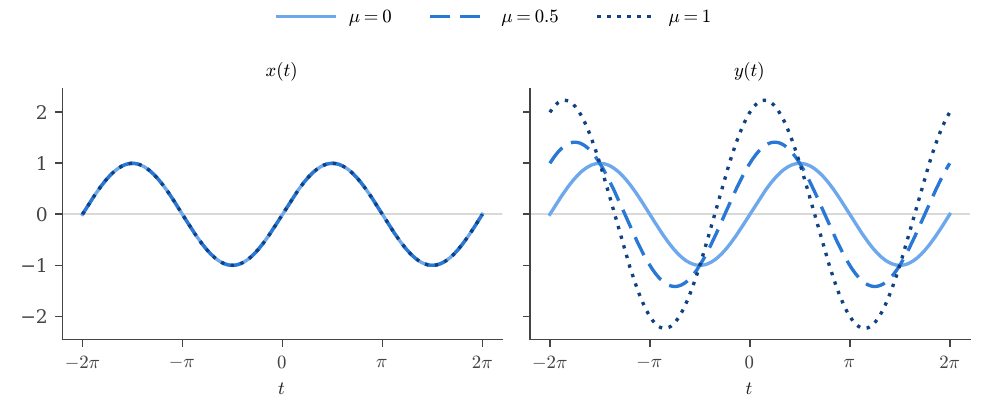}
\caption{$\PT$-symmetric solutions of the doubled system for $\lambda=-1/2$, $C=\tilde{C}=0$, $A=\tilde{A}=1$ and several values of $\mu$. The coordinate $x(t)=\sin t$ does not depend on $\mu$ (left), while the auxiliary coordinate $y(t)=\sin t+2\mu\cos t$ does (right).}
\label{fig:adjoint}
\end{figure}

Explicitly, by Section~\ref{sec:similarity}, $y=\Uop^\dagger y_0$, where $y_0$ is the solution~\eqref{eq:solution} of~\eqref{eq:mu0} with constants $\tilde C$ and $\tilde A$, and $\Uop^\dagger$ is given by~\eqref{eq:Udag}. This gives (Figure~\ref{fig:adjoint})
\begin{equation}
y(t)=
\begin{cases}
\tilde{C}+\tilde{A}\sinh(\sqrt{2\lambda}\,t)-\dfrac{2\mu\tilde{A}}{\sqrt{2\lambda}}\cosh(\sqrt{2\lambda}\,t), & \lambda>0,\\[2mm]
\tilde{C}+\tilde{A}t-\mu\tilde{A}t^2, & \lambda=0,\\[1mm]
\tilde{C}+\tilde{A}\sin(\sqrt{-2\lambda}\,t)+\dfrac{2\mu\tilde{A}}{\sqrt{-2\lambda}}\cos(\sqrt{-2\lambda}\,t), & \lambda<0.
\end{cases}
\label{eq:ysol}
\end{equation}

\subsection{The Hamiltonian}\label{sec:H}

To write $\Uop$ as a canonical transformation, we need the Hamiltonian of the doubled system. The canonical momenta of~\eqref{eq:L-doubled} are
\begin{equation}
\begin{aligned}
  P_1&=\dot y_1-\frac{\mu}{2}(y_2-y_1), & P_2&=\dot y_2-\frac{\mu}{2}(y_2-y_1),\\
  \pi_1&=\dot x_1-\frac{\mu}{2}(x_1+x_2), & \pi_2&=\dot x_2+\frac{\mu}{2}(x_1+x_2),
\end{aligned}
\label{eq:momenta}
\end{equation}
where $P_k$ is conjugate to $x_k$ and $\pi_k$ to $y_k$. The velocity Hessian is nonsingular, so there are no constraints, and the Legendre transform gives
\begin{equation}
  H=P_1\pi_1+P_2\pi_2+\frac{\mu}{2}(x_1+x_2)(P_1-P_2)+\frac{\mu}{2}(y_2-y_1)(\pi_1+\pi_2)-\lambda(x_1-x_2)(y_1-y_2).
  \label{eq:H}
\end{equation}
In terms of velocities, $H=\dot x_1\dot y_1+\dot x_2\dot y_2-\lambda(x_1-x_2)(y_1-y_2)$: the terms proportional to $\mu$ are linear in the velocities, so they do not contribute to the energy. On $\PT$-symmetric solutions $H=2\dot x(0)\dot y(0)$.

In the canonical variables
\begin{equation}
\begin{aligned}
  u_x&=x_1+x_2, & v_x&=x_1-x_2, & u_y&=y_1+y_2, & v_y&=y_1-y_2,\\
  P_{u_x}&=\tfrac12(P_1+P_2), & P_{v_x}&=\tfrac12(P_1-P_2), & P_{u_y}&=\tfrac12(\pi_1+\pi_2), & P_{v_y}&=\tfrac12(\pi_1-\pi_2),
\end{aligned}
\label{eq:CT1}
\end{equation}
the Hamiltonian becomes
\begin{equation}
  H=2\bigl(P_{u_x}P_{u_y}+P_{v_x}P_{v_y}\bigr)+\mu\,u_xP_{v_x}-\mu\,v_yP_{u_y}-\lambda\,v_xv_y ,
  \label{eq:H1}
\end{equation}
with the Hamilton equations
\begin{equation}
\begin{aligned}
  \dot u_x&=2P_{u_y}, & \dot v_x&=2P_{v_y}+\mu u_x, & \dot u_y&=2P_{u_x}-\mu v_y, & \dot v_y&=2P_{v_x},\\
  \dot P_{u_x}&=-\mu P_{v_x}, & \dot P_{v_x}&=\lambda v_y, & \dot P_{u_y}&=0, & \dot P_{v_y}&=\mu P_{u_y}+\lambda v_x .
\end{aligned}
\label{eq:Hamilton1}
\end{equation}
The canonical momentum $P_{u_y}$, conjugate to $u_y$, is conserved. It equals $\frac12\dot u_x$, half the mechanical total momentum $\pcm$ of the pair; the momentum conjugate to $u_x$ is $P_{u_x}=\frac12\dot u_y+\frac{\mu}{2}v_y$.

\subsection{Removing \texorpdfstring{$\mu$}{mu}}\label{sec:removing}

The characteristic polynomial of the linear system~\eqref{eq:Hamilton1} is $s^4(s^2-2\lambda)^2$ for every $\mu$. Indeed,~\eqref{eq:Hamilton1} splits into equations for $(u_x,P_{u_y},v_x,P_{v_y})$ and for $(v_y,P_{v_x},u_y,P_{u_x})$. In each set the first pair drives the second through $\mu$, but not conversely, so the matrix is block triangular and $\mu$ drops out of the determinant.

For $\lambda\neq0$ more is true: $\mu$ enters only through a change of variables, as in Section~\ref{sec:similarity}. By~\eqref{eq:similarity}, and since $(\Uop^{-1})^\dagger=(\Uop^\dagger)^{-1}$, the Bateman action can be written as
\begin{equation}
  -\int_{-T}^{T}y\,\E x\,\dd t=-\int_{-T}^{T}\tilde y\,\E_0\tilde x\,\dd t,\qquad
  \tilde x=\Uop x,\qquad \tilde y=(\Uop^\dagger)^{-1}y,
  \label{eq:action-tilde}
\end{equation}
up to boundary terms. So in the variables $\tilde x$ and $\tilde y$ the doubled system is the doubled system with $\mu=0$. It remains to write this change of variables in the canonical variables~\eqref{eq:CT1}.

For $\lambda\neq0$,~\eqref{eq:U-explicit} gives $\tilde x=x+\frac{\mu}{2\lambda}\dot u$, where $\dot u$ is odd. Similarly, $(\Uop^\dagger)^{-1}=1+\frac{\mu}{2\lambda}D(1-R)$ gives $\tilde y=y+\frac{\mu}{2\lambda}\dot v_y$, where $\dot v_y$ is even. Hence $\tilde u_x=u_x$, $\tilde v_x=v_x+\frac{\mu}{\lambda}\dot u_x$, $\tilde u_y=u_y+\frac{\mu}{\lambda}\dot v_y$ and $\tilde v_y=v_y$. By~\eqref{eq:Hamilton1}, $\dot u_x=2P_{u_y}$ and $\dot v_y=2P_{v_x}$. Completed by the momenta, the transformation reads
\begin{equation}
\begin{aligned}
  \tilde u_x&=u_x, & \tilde v_x&=v_x+\frac{2\mu}{\lambda}P_{u_y}, & \tilde u_y&=u_y+\frac{2\mu}{\lambda}P_{v_x}, & \tilde v_y&=v_y,\\
  \tilde P_{u_x}&=P_{u_x}+\frac{\mu}{2}v_y, & \tilde P_{v_x}&=P_{v_x}, & \tilde P_{u_y}&=P_{u_y}, & \tilde P_{v_y}&=P_{v_y}+\frac{\mu}{2}u_x .
\end{aligned}
\label{eq:CT2}
\end{equation}
It is generated by
\begin{equation}
  \Gen=\frac{2\mu}{\lambda}\,P_{v_x}P_{u_y}-\frac{\mu}{2}\,u_xv_y
  \label{eq:generator}
\end{equation}
in the sense that~\eqref{eq:CT2} is the flow of the Hamiltonian vector field of $\Gen$ at unit parameter, $\tilde z=z+\{z,\Gen\}+\frac12\{\{z,\Gen\},\Gen\}+\dots$ for each canonical variable $z$. The series stops after the first bracket, because $\{z,\Gen\}$ contains only $u_x$, $v_y$, $P_{u_y}$ and $P_{v_x}$, whose brackets with $\Gen$ vanish. Hence~\eqref{eq:CT2} is a canonical transformation. In the new variables the Hamiltonian~\eqref{eq:H1} becomes
\begin{equation}
  H=2\bigl(\tilde P_{u_x}\tilde P_{u_y}+\tilde P_{v_x}\tilde P_{v_y}\bigr)-\lambda\,\tilde v_x\tilde v_y ,
  \label{eq:H2}
\end{equation}
which is~\eqref{eq:H1} at $\mu=0$. Its Hamilton equations give $\ddot{\tilde u}_x=\ddot{\tilde u}_y=0$, $\ddot{\tilde v}_x=2\lambda\tilde v_x$ and $\ddot{\tilde v}_y=2\lambda\tilde v_y$.

Now $\mu$ enters only the map between the new and the physical variables. Since $\dot u_x=\pcm$,~\eqref{eq:CT2} gives
\begin{equation}
  v_x=\tilde v_x-\frac{\mu}{\lambda}\,\pcm .
  \label{eq:where-mu}
\end{equation}
This is the shift of Section~\ref{sec:model-dimer}, which vanishes for $\PT$-symmetric solutions.

For $\lambda=0$ the map~\eqref{eq:CT2} is singular, and in fact no time-independent linear change of variables can remove $\mu$. Such a change maps solutions to solutions and preserves the highest power of $t$ in them. For $\lambda=\mu=0$ every solution of~\eqref{eq:Hamilton1} is at most linear in $t$, whereas for $\mu\neq0$ the term $\mu\pcm t^2$ appears in $v_x$ (Section~\ref{sec:jordan-model}) and a term proportional to $\mu t^2$ in $u_y$, as in~\eqref{eq:ysol}. Correspondingly, at $\lambda=0$ the doubled system contains the Jordan blocks of Section~\ref{sec:jordan-model} twice, once for $x$ and once for $y$.

For $\lambda=0$, $\mu$ is removed instead by the time-dependent transformation $\Uop=1-\frac{\mu}{2}t(1+R)$. In the folded variables it gives $\tilde v_x=v_x-\mu t\,u_x$, the transformation to an accelerated frame. It is completed to a time-dependent canonical transformation, and the new Hamiltonian describes free motion and does not contain $\mu$ (Appendix~\ref{app:lambda0}).

%======================================================================
\section{The general linear equation}\label{sec:family}

We return to the general equation~\eqref{eq:general-linear}, whose solutions were obtained in~\cite{Mandal2025}; see~\cite{Yuan2025} for the higher-order equations $x^{(n)}(t)=\alpha x(t)+\beta x(-t)+f(t)$. By Section~\ref{sec:reduction}, it has a unique solution for any $x(0)$ and $\dot x(0)$, although at other initial times the Cauchy problem can fail to be uniquely solvable (footnote~\ref{fn:cauchy} and~\cite{Mandal2025}). We ask whether its solutions determine its coefficients.

Two equations of the form~\eqref{eq:general-linear} have the same solutions if and only if the difference of their coefficients, $(k_1,\dots,k_4)$, satisfies
\begin{equation*}
  k_1x(t)+k_2x(-t)+k_3\dot x(t)+k_4\dot x(-t)=0
\end{equation*}
for every solution of one of them, since both solution sets are two-dimensional. With $u=(1+R)x$ and $v=(1-R)x$, the even and odd parts of this condition are
\begin{equation}
  (k_1+k_2)\,u+(k_3+k_4)\,\dot v=0,\qquad (k_1-k_2)\,v+(k_3-k_4)\,\dot u=0.
  \label{eq:kconds}
\end{equation}
By Section~\ref{sec:reduction}, $u(0)$ and $\dot v(0)$ are arbitrary, while $\dot u(0)=v(0)=0$. At $t=0$ the first relation therefore gives $k_1+k_2=k_3+k_4=0$. The second relation vanishes at $t=0$, and its derivative there, with $\ddot u(0)=(a+b)u(0)+(c+d)\dot v(0)$ from~\eqref{eq:modes}, gives $(k_3-k_4)(a+b)=0$ and $k_1-k_2+(k_3-k_4)(c+d)=0$. So for $a+b\neq0$ all $k_i$ vanish, and the coefficients are uniquely determined by the solutions. On the surface $a+b=0$, which contains the model (Section~\ref{sec:springs}), the conditions leave the one-parameter family $(k_1,k_2,k_3,k_4)=\theta\,(\rho,-\rho,-1,1)$, with $\rho=c+d$. We now show that the equations related by the shift
\begin{equation}
  (a,b,c,d)\to(a,b,c,d)+\theta\,(\rho,-\rho,-1,1)
  \label{eq:shift}
\end{equation}
indeed have the same solutions.

For $b=-a$, the first equation of~\eqref{eq:modes} becomes $\ddot u=\rho\dot v$. So $\dot u-\rho v$ is conserved. For the model $\rho=0$, and this is the total momentum, which produces the Jordan block at $s=0$ (Section~\ref{sec:springs}). The conserved quantity vanishes for the solutions of~\eqref{eq:general-linear}, because $\dot u$ and $v$ are odd. Substituting $\dot u=\rho v$ into the second equation, we obtain
\begin{equation}
  \ddot v=\kappa v,\qquad \kappa=2a+c^2-d^2,
  \label{eq:kappa}
\end{equation}
and the general solution
\begin{equation}
  x(t)=C+A\bigl(\varphi(t)+\rho\,\psi(t)\bigr),\qquad \psi(t)=\int_0^t\varphi(t')\,\dd t',
  \label{eq:family-sol}
\end{equation}
where $\varphi$ is the odd solution of $\ddot\varphi=\kappa\varphi$ with $\dot\varphi(0)=1$. This agrees with the solutions obtained in~\cite{Mandal2025}. The solutions depend on $(a,c,d)$ only through $\rho$ and $\kappa$, and the shift~\eqref{eq:shift} leaves both unchanged:
\begin{equation*}
  \rho\to(c-\theta)+(d+\theta)=\rho,\qquad \kappa\to2(a+\theta\rho)+(c-\theta)^2-(d+\theta)^2=\kappa .
\end{equation*}
Choosing $\theta=c$, the equation has the same solutions as
\begin{equation}
  \ddot x(t)=a'\bigl[x(t)-x(-t)\bigr]+\rho\,\dot x(-t),\qquad a'=a+c(c+d),
  \label{eq:family-equiv}
\end{equation}
which has $c=0$ and therefore has a regular first-order Lagrangian (Section~\ref{sec:autonomous}). So the obstruction of Section~\ref{sec:EP-theorem} is a property of the equation and its local pair, not of the set of solutions. By Section~\ref{sec:notation}, $\E^\dagger$ is the operator of~\eqref{eq:general-linear} with $c\to-c$. So the adjoint equation has the same $\kappa$ but $\rho^\dagger=d-c$, and its solutions are $y=\tilde C+\tilde A(\varphi+\rho^\dagger\psi)$.

For~\eqref{eq:model} we have $\rho=0$, $\kappa=2\lambda$, $a'=\lambda$ and $\rho^\dagger=-2\mu$. The shift $(0,0,-1,1)$ changes only $\mu$, and~\eqref{eq:family-equiv} becomes~\eqref{eq:mu0}, the equation of the Lagrangian~\eqref{eq:Lx}. Equation~\eqref{eq:family-sol} reproduces~\eqref{eq:solution} and, up to the normalization of the constants,~\eqref{eq:ysol}.

Finally, a change of the velocity couplings alone, $(a,b,c,d)\to(a,b,c-\theta,d+\theta)$, leaves the solutions unchanged if and only if it is a move along~\eqref{eq:shift}, that is, if $a+b=0$ and $\rho=c+d=0$. For $a+b\neq0$ the solutions determine all coefficients, and for $\rho\neq0$ such a change shifts $\kappa$ by $-2\theta\rho$. These are the two exceptional-point conditions of Sections~\ref{sec:velocity} and~\ref{sec:springs}, which together define the model~\eqref{eq:model}. Being at an exceptional point is not enough, however: the direction of the one-way coupling matters. On the other exceptional line $c=d$, the relative coordinate drives the total momentum, and a change of the velocity couplings does change the solutions. The adjoint equation is an example: its velocity matrix lies on $c=d$ (Section~\ref{sec:bateman}), and its solutions depend on $\mu$ through $\rho^\dagger=-2\mu$.

%======================================================================
\section{Conclusion}\label{sec:conclusion}

We have shown that every second-order differential equation with time reflection describes the $\PT$-symmetric solutions of a local pair of oscillators that is invariant under exchange combined with time reversal. The nonlocality only reflects this symmetry together with symmetric initial data. For linear equations the pair contains only springs, balanced gain and loss, and a gyroscopic coupling. Such a pair has a regular first-order Lagrangian unless its oscillators are coupled by a spring and its velocity matrix is a single Jordan block, which is Case~IIIb of Douglas's classification~\cite{Douglas1941}, and the solutions of the equation determine its coefficients unless there is no on-site spring.

For equation~\eqref{eq:model}, $\mu$ multiplies only the total momentum of the pair, which vanishes on $\PT$-symmetric solutions. In operator form, $\E=\Uop^{-1}\E_0\Uop$, where $\Uop$ differs from the identity by a nilpotent operator. Among the linear equations with reflection, the model is singled out by the conditions $c+d=0$ and $a+b=0$. Each gives a Jordan block of the pair and is an exceptional point of the corresponding larger family, and together they are exactly the conditions under which a change of the velocity couplings leaves the solutions unchanged. The direction of the one-way coupling also matters: on the other exceptional line $c=d$, as for the adjoint equation, such a change does change the solutions.

The same structure decides the variational questions. For $\lambda\mu\neq0$ the pair of the model has no regular first-order Lagrangian, with any multiplier. The substitution $x\to\Uop x$ gives instead a Lagrangian with second derivatives for $\lambda\neq0$, and a first-order Lagrangian with time-dependent coefficients for $\lambda=0$, so an auxiliary variable is not necessary. In the Bateman doubling, the change of variables $(x,y)\to(\Uop x,(\Uop^\dagger)^{-1}y)$ removes $\mu$ from the Hamiltonian: by a time-independent canonical transformation for $\lambda\neq0$, and by the transformation to an accelerated frame for $\lambda=0$.

By the reduction of Section~\ref{sec:reduction}, $\PT$-symmetric pairs started with $\PT$-symmetric data realize differential equations with reflection~\cite{WienerAftabizadeh1985,CabadaTojo2015}, as the active $LRC$ dimers of~\cite{Schindler2011,Schindler2012} and the circuit of Section~\ref{sec:circuit} show. Conversely, results on equations with involutions, such as Green's functions and maximum principles~\cite{CabadaTojo2015}, give statements about $\PT$-symmetric solutions.

There are several directions to extend our work. It would be interesting to quantize the doubled Hamiltonian~\eqref{eq:H2}, as was done for Bateman's oscillator~\cite{FeshbachTikochinsky1977,Celeghini1992,BlasoneJizba2004,Valdez2025,Beau2026} and for $\PT$-symmetric pairs~\cite{Bender2013PRA}. It would also be interesting to study nonlinear pairs. For example, the dimer of~\cite{Cuevas2013} gives the nonlinear equation with reflection $\ddot x(t)=-x(t)+k\,x(-t)+\gamma\dot x(t)+\epsilon\, x(t)^3$, in their notation except that we write $k$ for their coupling. Equations with reflection of higher order~\cite{Yuan2025} correspond in the same way to pairs with higher derivatives, and their variational structure remains to be understood. Finally, one could study reverse-time nonlocal field equations, which have the same symmetry~\cite{AblowitzMusslimani2017} or the same symmetry combined with complex conjugation~\cite{Yang2018}, extending the spatially nonlocal formalism of~\cite{Pazarci2026}. We hope to return to these questions elsewhere.

\section*{Acknowledgements}

Ilmar Gahramanov is supported by the Istanbul Integrability and Stringy Topics Initiative (\href{https://istringy.org}{istringy.org}). Both authors are supported by the Bo\u{g}azi\c{c}i University Research Fund under grant number Y26.

\section*{Data availability statement}

All data that support the findings of this study are included within the article.

%======================================================================
\appendix

\section{The pair with balanced gain and loss}\label{sec:bender}

For $d=0$ the system~\eqref{eq:dimer} is the $\PT$-symmetric pair of~\cite{Bender2013PRA},
\begin{equation*}
  \ddot x_1+\omega^2x_1+2\gamma\dot x_1=-\epsilon x_2,\qquad
  \ddot x_2+\omega^2x_2-2\gamma\dot x_2=-\epsilon x_1,
\end{equation*}
with $a=-\omega^2$, $b=-\epsilon$ and $c=-2\gamma$. Its $\PT$-symmetric solutions satisfy $x_2(t)=x_1(-t)$, and $x\equiv x_1$ obeys
\begin{equation}
  \ddot x(t)+\omega^2x(t)+2\gamma\dot x(t)+\epsilon\, x(-t)=0.
  \label{eq:bender}
\end{equation}
We take $\gamma>0$ and $\epsilon>0$; the other signs follow by exchanging the two oscillators or by $x_2\to-x_2$. For $\omega^2>2\gamma^2$ the pair has an unbroken $\PT$ phase, in which all frequencies are real, for $2\gamma\sqrt{\omega^2-\gamma^2}<\epsilon<\omega^2$~\cite{Bender2013PRA}; for $\omega^2\le2\gamma^2$ it has none.\footnote{The frequencies are $\Omega^2=\omega^2-2\gamma^2\pm\sqrt{\epsilon^2-4\gamma^2\omega^2+4\gamma^4}$.} By Section~\ref{sec:linear}, these $\PT$ phases carry over to~\eqref{eq:bender}. At $\epsilon=\omega^2$, for example, the exponent $s=0$ has $K\bm\xi=0$ with $\bm\xi=(1,-1)$ and $\sigma_x\bm\xi=-\bm\xi$, and the corresponding $\PT$-symmetric solution of~\eqref{eq:bender} is $x(t)=t-\gamma/\omega^2$.

The active $LRC$ dimers of~\cite{Schindler2011,Schindler2012} obey systems of the form~\eqref{eq:dimer} in the capacitor charges, with the operation $\sigma$ of Section~\ref{sec:reduction}. For inductive coupling $d=0$, and they are the pair above.\footnote{For coupling through a capacitor $C_c$, the node fluxes obey a system with the mass matrix $(C+C_c)\bigl(\begin{smallmatrix}1&\eta\\ \eta&1\end{smallmatrix}\bigr)$, with $\eta=-C_c/(C+C_c)$. For $\PT$-symmetric solutions this gives $\ddot x(t)+\eta\ddot x(-t)=a\,x(t)+c\,\dot x(t)$. Since $(1+\eta R)^{-1}=(1-\eta R)/(1-\eta^2)$, this equation has the form~\eqref{eq:general-linear}, with the coefficient $-\eta c/(1-\eta^2)$ of $\dot x(-t)$. So a capacitively coupled dimer also contains a gyroscopic term~\cite{Schindler2012}.} Started with $\PT$-symmetric data, such a dimer gives one solution of~\eqref{eq:bender} on the whole line: $x_1(t)$ for $t\ge0$ and $x_2(-t)$ for $t\le0$. For generic data, such as a current injected into one circuit~\cite{Schindler2012}, a record is the sum of a solution of~\eqref{eq:bender} and a solution of the equation obtained from $\sigma'$-invariant data, which is~\eqref{eq:bender} with $\epsilon\to-\epsilon$. The pendula of~\cite{Bender2013AJP} and the microcavities of~\cite{Peng2014} obey~\eqref{eq:bender} only as models: in~\cite{Bender2013AJP} gain and loss are applied as timed impulses, and in~\cite{Peng2014} they are only nearly balanced.

\section{The Lagrangian with second derivatives}\label{app:second-order}

The Euler--Lagrange equation of~\eqref{eq:L2} follows most easily from the variation of the action. Since the Euler--Lagrange equation of~\eqref{eq:Lx} is $-2\E_0x=0$, the variation of the action at $\tilde x$ is $\delta S=-2\int(\E_0\tilde x)\,\delta\tilde x\,\dd t$, with $\delta\tilde x=\Uop\,\delta x$. We take $\delta x$ and $\delta\dot x$ to vanish at $t=\pm T$ and move $\Uop$ to the other factor. With~\eqref{eq:intertwine}, this gives
\begin{equation}
  \delta S=-2\int\bigl(\Uop^\dagger\E_0\Uop x\bigr)\,\delta x\,\dd t=-2\int\bigl(\Uop^\dagger\Uop\,\E x\bigr)\,\delta x\,\dd t .
  \label{eq:deltaS}
\end{equation}
So the Euler--Lagrange equation is $\Uop^\dagger\Uop\,\E x=0$, where, by the identities of Section~\ref{sec:notation},
\begin{equation*}
  \Uop^\dagger\Uop=1+\frac{\mu}{\lambda}DR-\frac{\mu^2}{2\lambda^2}D^2(1+R).
\end{equation*}
This equation contains the fourth derivative of $u$. For a fourth-order equation one would normally expect two additional initial conditions, and solutions that do not solve~\eqref{eq:model}. This does not happen here, because $\Uop^\dagger\Uop$ has an inverse that is again a differential operator: $\Uop^{-1}$ is $\Uop$ with $\mu$ replaced by $-\mu$, and $(\Uop^\dagger)^{-1}=(\Uop^{-1})^\dagger$. Applying $\Uop^{-1}(\Uop^\dagger)^{-1}$ to $\Uop^\dagger\Uop\,\E x=0$ gives $\E x=0$ directly, without integration and hence without new constants. By contrast, an equation $D^2(\E x)=0$ would only give $\E x=A+Bt$. Thus the stationary points of~\eqref{eq:L2} are exactly the solutions of~\eqref{eq:model}, with the initial data of~\eqref{eq:model}.

In the folded form,~\eqref{eq:xtilde} reads $\tilde x_1=x_1+\beta\pcm$ and $\tilde x_2=x_2-\beta\pcm$. The folded Lagrangian is $\dot{\tilde x}_1^2+\dot{\tilde x}_2^2+\lambda(\tilde x_1-\tilde x_2)^2$, and its stationary points are exactly the solutions of the pair~\eqref{eq:local-model}, for all initial data. Still,~\eqref{eq:L2} is not a Lagrangian in the sense of Sections~\ref{sec:lagrangians} and~\ref{sec:nonlocal-L}. It depends on the accelerations only through $\ddot u=\dot p_{\mathrm{cm}}$, so its Hessian with respect to $(\ddot x_1,\ddot x_2)$ is singular. The Ostrogradsky momenta conjugate to $\dot x_1$ and $\dot x_2$ are therefore equal, and the Hamiltonian formulation has a primary constraint. The Euler--Lagrange equations are equivalent to the Hamilton--Dirac equations with this constraint~\cite{Pons1989}; see also~\cite{Coban2023}.

\section{The Bateman Lagrangian}\label{app:bateman}

In the original variables, the Lagrangian~\eqref{eq:L1-doubled} is $L_1=-y\,\E x+\dot F+\Omega$, where
\begin{equation}
\begin{aligned}
F&=y(t)\dot x(t)-\frac{\mu}{2}x(t)y(t)-\frac{\mu}{4}\bigl[x(-t)y(t)-x(t)y(-t)\bigr],\\
\Omega&=\frac{\lambda}{2}\bigl[x(-t)y(t)-x(t)y(-t)\bigr]+\frac{\mu}{4}\bigl[x(-t)\dot y(t)+\dot x(-t)y(t)-x(t)\dot y(-t)-\dot x(t)y(-t)\bigr].
\end{aligned}
\label{eq:F-Omega}
\end{equation}
The total derivative gives only boundary terms, and $\Omega$ is odd in $t$, so it integrates to zero on $[-T,T]$.

\section{Removing \texorpdfstring{$\mu$}{mu} for \texorpdfstring{$\lambda=0$}{lambda=0}}\label{app:lambda0}

For $\lambda=0$ the map~\eqref{eq:CT2} is singular, and we use instead $\Uop=1-\frac{\mu}{2}t(1+R)$ and $(\Uop^\dagger)^{-1}=1+\frac{\mu}{2}t(1-R)$. In the folded variables they give
\begin{equation}
  \tilde u_x=u_x,\qquad \tilde v_x=v_x-\mu t\,u_x,\qquad \tilde u_y=u_y+\mu t\,v_y,\qquad \tilde v_y=v_y .
  \label{eq:accelerated}
\end{equation}
The second relation is the transformation to an accelerated frame. Indeed, $\ddot v_x=2\mu\pcm$ for $\lambda=0$. Since $\dot u_x=\pcm$ is constant, $\ddot{\tilde v}_x=\ddot v_x-2\mu\dot u_x=0$. In the same way the auxiliary coordinates obey $\ddot v_y=0$ and $\ddot u_y=-2\mu\dot v_y$, and $\tilde u_y$ moves freely.

As for $\lambda\neq0$ in Section~\ref{sec:removing}, the transformation is completed by the momenta with a generator, which now depends on time:
\begin{equation}
  \Gen_0=-\mu t\,\bigl(u_xP_{v_x}-v_yP_{u_y}\bigr).
  \label{eq:generator0}
\end{equation}
At fixed $t$, the nonzero brackets with $\Gen_0$ are $\{v_x,\Gen_0\}=-\mu t\,u_x$, $\{u_y,\Gen_0\}=\mu t\,v_y$, $\{P_{u_x},\Gen_0\}=\mu t\,P_{v_x}$ and $\{P_{v_y},\Gen_0\}=-\mu t\,P_{u_y}$. They contain only variables with vanishing brackets with $\Gen_0$, so again the flow of $\Gen_0$ at unit parameter is $\tilde z=z+\{z,\Gen_0\}$, a canonical transformation at each $t$. It gives~\eqref{eq:accelerated} together with $\tilde P_{u_x}=P_{u_x}+\mu t\,P_{v_x}$ and $\tilde P_{v_y}=P_{v_y}-\mu t\,P_{u_y}$, and leaves the other momenta unchanged.

Since the transformation depends on time, the new Hamiltonian is not simply $H$ in the new variables. With $\Gen_0=t\,\Gen_1$, the map is the flow of $\Gen_1$ at parameter $t$, so $\partial\tilde z/\partial t=\{\tilde z,\Gen_1\}$, and along the motion
\begin{equation*}
  \frac{\dd\tilde z}{\dd t}=\{\tilde z,H\}+\frac{\partial\tilde z}{\partial t}=\Bigl\{\tilde z,\,H+\frac{\partial\Gen_0}{\partial t}\Bigr\}.
\end{equation*}
So the new variables obey Hamilton's equations with the Hamiltonian $H+\partial\Gen_0/\partial t$. Here $\partial\Gen_0/\partial t=-\mu\,(u_xP_{v_x}-v_yP_{u_y})$ cancels the terms proportional to $\mu$ in~\eqref{eq:H1}, and $2\bigl(P_{u_x}P_{u_y}+P_{v_x}P_{v_y}\bigr)=2\bigl(\tilde P_{u_x}\tilde P_{u_y}+\tilde P_{v_x}\tilde P_{v_y}\bigr)$. Hence the new Hamiltonian,
\begin{equation}
  \tilde H=H+\frac{\partial\Gen_0}{\partial t}=2\bigl(\tilde P_{u_x}\tilde P_{u_y}+\tilde P_{v_x}\tilde P_{v_y}\bigr),
  \label{eq:H3}
\end{equation}
describes free motion and does not contain $\mu$.

\bibliographystyle{utphys}
\bibliography{references}

\end{document}